\documentclass[11pt]{article}
\usepackage{geometry}

\usepackage{graphics,graphicx,color,float, hyperref}
\usepackage{epsfig,bbm}
\usepackage{bm}
\usepackage{mathtools}
\usepackage{amsmath}
\usepackage{amssymb}
\usepackage{amsfonts}
\usepackage{amsthm}
\usepackage {times}
\usepackage{layout}
\usepackage{setspace}

\newtheorem*{lemma*}{Lemma}

\newtheorem{fact}{Fact}

\newcommand{\beq}{\begin{equation}}
\newcommand{\enq}{\end{equation}}
\newcommand{\bel}{\begin{lemma}}
\newcommand{\enl}{\end{lemma}}
\newcommand{\bet}{\begin{theorem}}
\newcommand{\ent}{\end{theorem}}

\newcommand{\tr}{\mathrm{Tr}}

\newcommand{\Tr}{\mathrm{Tr}}
\newcommand{\ketbra}[1]{|#1\rangle\langle#1|}

\newcommand{\eps}{\varepsilon}

\newcommand {\dmineps} [3] {\fn{\mathrm{D}^{#3}_{\mathrm{H}}}{#1 \middle\| #2}}

\newcommand*{\cH}{\mathcal{H}}

\newcommand*{\cD}{\mathcal{D}}

\newcommand*{\cN}{\mathcal{N}}

\newcommand*{\cX}{\mathcal{X}}

\newcommand{\cL}{\mathcal{L}}

\newcommand{\suppress}[1]{}

\newcommand{\defeq}{\ensuremath{ \stackrel{\mathrm{def}}{=} }}
\newcommand{\F}{\mathrm{F}}
\newcommand{\Pur}{\mathrm{P}}
\newcommand {\br} [1] {\ensuremath{ \left( #1 \right) }}

\newcommand {\minusspace} {\: \! \!}
\newcommand {\smallspace} {\: \!}
\newcommand {\fn} [2] {\ensuremath{ #1 \minusspace \br{ #2 } }}

\newcommand {\relent} [2] {\fn{\mathrm{D}}{#1 \middle\| #2}}
\newcommand {\varent} [2] {\fn{\mathrm{V}}{#1 \middle\| #2}}
\newcommand {\dmax} [2] {\fn{\mathrm{D}_{\max}}{#1 \middle\| #2}}
\newcommand {\mutinf} [2] {\fn{\mathrm{I}}{#1 \smallspace : \smallspace #2}}
\newcommand {\imax}{\ensuremath{\mathrm{I}_{\max}}}

\newcommand {\hmineps} [2] {\fn{\mathrm{H}^{#2}_{\min}}{#1}}

\newcommand {\hmaxeps} [3] {\fn{\mathrm{H}^{#3}_{\max}}{#1 \middle | #2}}
\newcommand {\dheps} [3] {\ensuremath{\mathrm{D}_{\mathrm{H}}^{#3}\left(#1 \| #2\right)}}
\newcommand {\id} {\ensuremath{\mathbb{I}}}
\newcommand {\bbeta} {\ensuremath{\bar{\beta}}}

\newcommand*{\cY}{\mathcal{Y}}

\newcommand{\bra}[1]{\langle #1|}
\newcommand{\ket}[1]{|#1 \rangle}

\mathchardef\mhyphen="2D

\newcommand*{\Iheps}[3]{\ensuremath{\mathrm{I}^{#3}_{\mathrm{H}}\left(#1:#2 \right)}}
\newcommand*{\Ihepst}[3]{\ensuremath{\mathrm{\tilde{I}}^{#3}_{\mathrm{H}}\left(#1:#2 \right)}}
\newcommand*{\Ihepsh}[3]{\ensuremath{\mathrm{\hat{I}}^{#3}_{\mathrm{H}}\left(#1:#2 \right)}}
\newcommand*{\Ihepstres}[4]{\ensuremath{\mathrm{\tilde{I}}^{#3, #4}_{\mathrm{H}}\left(#1:#2 \right)}}
\newcommand*{\Ihepshres}[4]{\ensuremath{\mathrm{\hat{I}}^{#3, #4}_{\mathrm{H}}\left(#1:#2 \right)}}

\newcommand*{\conv}{\mathrm{conv}}

\makeatletter
\newcommand*{\rom}[1]{\expandafter\@slowromancap\romannumeral #1@}
\makeatother

\mathchardef\mhyphen="2D

\newtheorem{definition}{Definition}
\newtheorem{claim}{Claim}

\newtheorem{theorem}{Theorem}
\newtheorem{lemma}{Lemma}
\newtheorem{corollary}{Corollary}
\begin {document}
\title{A hypothesis testing approach for communication over entanglement assisted compound quantum channel}
\author{
Anurag Anshu\footnote{Centre for Quantum Technologies, National University of Singapore, Singapore. \texttt{a0109169@u.nus.edu}} \qquad
Rahul Jain\footnote{Centre for Quantum Technologies, National University of Singapore and MajuLab, UMI 3654, 
Singapore. \texttt{rahul@comp.nus.edu.sg}} \qquad 
Naqueeb Ahmad Warsi\footnote{Centre for Quantum Technologies, National University of Singapore and School of Physical and Mathematical Sciences, Nanyang Technological University, Singapore and IIITD, Delhi. \texttt{warsi.naqueeb@gmail.com}} 
}
\date{}
\maketitle

\abstract{

We study the problem of communication over a compound quantum channel in the presence of entanglement. Classically such channels are modeled as a collection of conditional probability distributions wherein neither the sender nor the receiver is aware of the channel being used for transmission, except for the fact that it belongs to this collection. We provide near optimal achievability and converse bounds for this problem in the one-shot quantum setting in terms of quantum hypothesis testing divergence. We also consider the case of informed sender, showing a one-shot achievability result that converges appropriately in the asymptotic and i.i.d. setting. Our achievability proof is similar in spirit to its classical counterpart. To arrive at our result, we use the technique of \textit{position-based decoding} along with a new approach for constructing a \textit{union} of two projectors, which can be of independent interest. We give another application of the union of projectors to the problem of testing composite quantum hypotheses.  

 }

\section{Introduction}

A typical assumption while communicating over a channel is that the communicating parties are aware of the channel characteristics. This means that the parties know the output distribution (or the quantum state) for a given input (or a quantum state). This leads to the well known model of point to point channel, which has been extensively studied in literature starting from the seminal work of Shannon \cite{Shannon}. In this model, a natural problem is to find the number of messages that can be communicated through a channel (or a quantum channel if quantum communication is considered) such that the probability of incorrectly decoding a message is upper bounded by a fixed error parameter $\eps$. In the asymptotic and i.i.d. (independent and identically distributed) setting, where the communicating parties are allowed many independent uses of the channel, the number of bits that can be transmitted per channel use (for arbitrarily small error) is called the capacity of the channel.

It is not hard to imagine a real world setting which differs from this point of view. Let us consider the case where the communicating parties are not completely aware of the channel characteristics, potentially due to the lack of sufficient statistical data or chaotic behavior of the channel. In such a case, the parties may have to work with the assumption that the channel is an element of a finite collection of channels. This setting is known as the compound channel. It has a straightforward quantum version where the channel is an element of a finite collection of quantum maps $\{\cN^{(1)},\cN^{(2)},\ldots, \cN^{(s)}\}$. In this work, we consider the entanglement assisted classical communication over such compound quantum channels in the one-shot setting. In contrast to the asymptotic and i.i.d.  setting, the one-shot setting concerns with a single use of the channel and provides a natural framework for communication over realistic classical or quantum networks. Furthermore, the results obtained in the one-shot setting can be used to recover the corresponding asymptotic and i.i.d. results by appropriate analysis of many independent instances of a channel. By standard duality between teleportation \cite{Teleportation93} and superdense coding \cite{bennett92}, our results also apply to the one-shot entanglement assisted quantum capacity. 

The problem of communication over the compound (classical) channel in the asymptotic and i.i.d. setting was studied in \cite{blackwell1959, Wolfowitz59} (see also \cite[Theorem 7.1]{GamalK12}). The quantum capacities of compound quantum channels have been studied in several works, such as \cite{BBoche09, Hayashi2009, Bjelakovic2013, Mosonyi15, BBocheN08, BBocheN09, Bjelakovic2009, BertaGW17, BocheJK2017}. The works \cite{BertaGW17, BocheJK2017} studied the entanglement assisted capacities in the asymptotic and i.i.d. setting, where optimal results were obtained in the asymptotic and i.i.d. setting. The key tool used in \cite{BertaGW17}, which also studied the one-shot version of this problem, was that of the decoupling method and their bounds were obtained in terms of smooth conditional min and max entropies. 

\vspace{0.1in}

\noindent {\bf Quantum hypothesis testing in quantum channel coding:} In recent years, the quantum hypothesis testing approach has served as a crucial tool for the analysis of quantum channel coding. In the asymmetric quantum hypothesis testing, two quantum states $\rho, \sigma$ are given, and one is required to construct a measurement $M$ that succeeds on $\sigma$ with as small probability as possible (measured by the Type $2$ error $\Tr(M\sigma)$), given that it succeeds on $\rho$ with probability close to $1$ (the probability of failure on $\rho$ is measured by the Type $1$ error $\Tr((\id - M)\rho)$). The Type $2$ error is captured by the quantum hypothesis testing divergence. Several applications of this approach to the entanglement unassisted quantum channel coding problem have been given in the asymptotic and non-i.i.d. setting in the works \cite{OgawaN00, Ogawa:2002, HyashiN03, Hayashi07, Mosonyi15} and in the one-shot setting, in the work \cite{WangR12}. For the entanglement assisted quantum channel coding, a converse bound based on the quantum hypothesis testing divergence has been shown in \cite{MatthewsW14}. This has been shown to be near optimal in the work \cite{AnshuJW17}, which gives a hypothesis testing based achievability result for the entanglement assisted quantum channel coding, by introducing the technique of position-based decoding. 

\vspace{0.1in}

\noindent {\bf Our results:} In this work, we give a one-shot achievability result for the task of entanglement assisted communication over compound quantum channel in terms of the quantum hypothesis testing divergence. Broadly, our technique follows the position-based decoding method, which allows for quantum hypothesis testing to be performed on several registers. But we require a new technical component along with that used in \cite{AnshuJW17}, that we discuss below. 

\vspace{0.1in}

\noindent \emph{A union of quantum projectors:} A key challenge that arises is to formulate a suitable quantum version of the union of two events in probability theory. In the classical achievability result for communication over compound channel, one uses a statement of the form $\Pr\left\{E_1\cup E_2\right\} \geq \max\{\Pr\left\{E_1\right\}, \Pr\left\{E_2\right\}\}$, for two events $E_1$ and $E_2$ (see for example, \cite[Theorem 7.1]{GamalK12}), which is a converse to the union bound that states $\Pr\left\{E_1\cup E_2\right\} \leq \Pr\left\{E_1\right\}+ \Pr\left\{E_2\right\}$. Quantum analogues of both these statements have been studied in previous works. Two well known examples of quantum version of the union bound are the Hayashi-Nagaoka inequality \cite{HyashiN03} and the results on sequential measurement \cite{Sen12, Wilde2013, Gao15}. The converse to the union bound has been studied in the quantum setting in \cite{Aaronson06} (where it is called ``Quantum OR bound'') and in \cite{HarrowLM17}. The result in \cite[Corollary 11]{HarrowLM17} is as follows. Consider a collection of projectors $\{\Pi_1,\Pi_2,\ldots \Pi_s\}$ and a quantum state $\rho$, with the property that either there exists an $i$ such that $\Tr[\Pi_i\rho]\geq 1-\eps$ or $\mathbb{E}_i \Tr[\Pi_i\rho] \leq \delta$. Then there exists a projector $\Pi$ with the property that in the first case, $\Tr[\Pi\rho] \geq \frac{(1-\eps)}{7}$ and in the second case, $\Tr[\Pi\rho] \leq 4\delta s$. 

The above result cannot be used for our purpose since we require a one-shot decoding strategy which makes an error of at most $\eps$ in probability, for every $\eps \in (0,1)$. To overcome this, we prove a new quantum analogue of the converse to the union bound in Lemma \ref{uni}, which is more suited for our application. This is a main technical contribution of this paper and uses Jordan's lemma (on the joint structure of two projectors) at its core. Informally, the statement of Lemma \ref{uni} is as follows. 
\begin{lemma*}[Informal]
Consider a collection of projectors $\{\Pi_1,\Pi_2,\ldots \Pi_s\}$ and quantum states $\{\rho_1,\rho_2,\ldots \rho_s\}$ such that for all $i$, $\Tr[\Pi_i\rho_i] \geq 1-\eps$. Then there exists a projector $\Pi^*$ such that it succeeds well on all $\rho_i$ (that is, $\Tr[\Pi^*\rho_i] \geq 1- O(\eps)$) and for all quantum states $\sigma$, $\Tr[\Pi^*\sigma] \leq s^{O(\log\frac{\log(s)}{\eps})}\mathbb{E}_i \Tr[\sigma\Pi_i]$. 
\end{lemma*}

\vspace{0.1in}

\noindent \emph{One-shot results for the entanglement assisted compound quantum channel:} Using Lemma \ref{uni} and the aforementioned position-based decoding, we give our achievability result in Theorem \ref{achievability}. Our achievability result is in terms of a one-shot quantity derived from the quantum hypothesis testing divergence. More precisely, let $\dheps{\rho}{\sigma}{\eps}$ denote the smooth quantum hypothesis testing divergence. We consider the following variant of the smooth quantum hypothesis testing divergence, for a bipartite quantum state $\rho_{AB}$: $$\Iheps{A}{B}{\eps}_{\rho_{AB}} := \min_{\sigma_A}\dheps{\rho_{AB}}{\sigma_A\otimes \rho_B}{\eps},$$ where the minimization is over all quantum states $\sigma_A$. This quantity appeared earlier in the work \cite{MatthewsW14} in context of converse bounds for the entanglement assisted classical communication over noisy quantum channels. Our achievability result in Theorem \ref{achievability} states that given $\left\{\cN^{(i)}_{A \to B }\right\}_{i=1}^s$ as a compound quantum channel that takes a register $A$ as input and outputs a register $B$, it is possible to communicate at least
$$\max _{\ketbra{\psi}_{A A'}} \left( \min_{i \in [1:s]} \Iheps{B}{A'}{\eps}_{\cN^{(i)}_{A \to B } (\ketbra{\psi}_{AA'})} - 2\log s \cdot\log \left(\frac{\log s}{\eta}\right) - 2\log\frac{s}{\eps}\right),$$
number of bits through the channel, with an error of at most $\eps + 3\eta$. Here $\eps, \eta$ are error parameters and the error is measured in terms of the probability of incorrectly decoding a message sent by Alice.

Using the techniques developed in \cite{MatthewsW14}, we also give a converse bound for our task (which was also observed in the asymptotic and i.i.d. setting in \cite{BertaGW17}) in terms of $\Iheps{A}{B}{\eps}_{\rho}$. The converse, appearing in Theorem \ref{theo:converse}, says that given $\left\{\cN^{(i)}_{A \to B }\right\}_{i=1}^s$ as a compound quantum channel, any entanglement assisted protocol that makes an error of $\eps$ can communicate at most 
$$\max _{\ketbra{\psi}_{A A'}} \left( \min_{i \in [1:s]} \Iheps{B}{A'}{\eps}_{\cN^{(i)}_{A \to B } (\ketbra{\psi}_{AA'})}\right)$$
number of bits through the channel. Thus, the achievability and converse bounds match up to an additive factor of $O(\log s\cdot\log\frac{\log s}{\eps})$. The additive loss of $O(\log\frac{s}{\eps})$ in the amount of communication can also be found in the one-shot classical case (for example, in the one-shot version of the argument given in \cite[Theorem 7.1]{GamalK12}). We also argue that an additive loss of $O(\log s)$ cannot be avoided in general.  

At this stage, we highlight the importance of the union bound (Lemma \ref{uni}) in our context. To obtain the achievability results for the compound quantum channel, the previous works \cite{BertaGW17, BBocheN08} considered the average quantum channel $\frac{1}{s}\sum_i\cN^{(i)}_{A \to B }$ and argued that an error of $\eps$ for the average quantum channel leads to an error at most $s\eps$ for each channel. This technique cannot work for proving Theorem \ref{achievability}, since the loss in error from $\eps$ to $s\eps$ gets reflected in the Type $1$ error of the quantum hypothesis testing. On the other hand, the application of Lemma \ref{uni} leads to a minor increase in the Type $1$ error from $\eps$ to $\eps+\delta$. An increase in error that depends on $s$ only appears in the Type $2$ error. This loss gets reflected in the number of bits that can be transmitted, leading to the additive loss of $O(\log s\log\log s)$. 

An important question is to establish the asymptotic and i.i.d. properties of our bound. For this, we relate the quantity $\Iheps{A}{B}{\eps}_{\rho_{AB}}$ to the quantity $\dheps{\rho_{AB}}{\rho_A\otimes\rho_B}{\eps}$ in Lemma \ref{theo:Ihdhsame}, where we use the converse result in \cite{MatthewsW14} and the achievability result in \cite{AnshuJW17}. The achievability result in the asymptotic and i.i.d. setting can then be obtained by appealing to the asymptotic and i.i.d. behavior of $\dheps{\rho_{AB}}{\rho_A\otimes\rho_B}{\eps}$ \cite{TomHay13, li2014}, which we show in Corollary \ref{cor:asympcomp}. A matching converse in the asymptotic and i.i.d. setting has been given in \cite{BertaGW17}, again using the ideas developed in \cite{MatthewsW14}. We note that these results can also be extended to the case of infinite compound quantum channel (where the number of channels in the set is infinite) by appropriate discretization argument, as has been discussed in details in \cite{BertaGW17, BBocheN08}.

\vspace{0.1in}

\noindent \emph{Compound quantum channel with informed sender:} We also consider a model of the compound quantum channel where the sender is aware of $s$, the label of the channel. In this setting, the sender knows which channel is acting from the given collection, but the receiver has no such information. This was considered in the classical asymptotic and i.i.d. setting in \cite{Wolfowitz78} and in the quantum one-shot and asymptotic and i.i.d. settings in \cite{BertaGW17}. We give our one-shot achievability result in terms of an another variant of the aforementioned one-shot version of quantum mutual information, which appears in Section \ref{sec:informedsender}. Our protocol closely follows the protocol used for the case of uninformed sender (Theorem \ref{achievability}). We provide the asymptotic and i.i.d. analysis in the same section, showing the convergence to the optimal rate in this setting (Theorem \ref{theo:asympinfo}). 

\vspace{0.1in}

\noindent \emph{Composite quantum hypothesis testing:} As another application of Lemma \ref{uni}, we consider the problem of composite quantum hypothesis testing, introduced in \cite{BertaBH17}. Let $S_1, S_2$ be two sets of quantum states on a register $A$ and for an integer $n>1$, let the sets $S^n_1, S^n_2$ be defined as follows. For $i\in \{1,2\}$, $S^n_i = \text{conv}(\rho^{\otimes n}: \rho\in S_i)$ (where `$\text{conv}$' refers to the convex hull). The problem is to design a positive operator $\Lambda_n \preceq \id$, such that $\Tr(\Lambda_n\rho) \geq 1-\eps$ (with $\eps\in (0,1)$) for all $\rho\in S^n_1$ and $\Tr(\Lambda_n\sigma)$ is as small as possible for all $\sigma\in S^n_2$. The case where the set $S_2$ is singleton was considered in \cite{Mosonyi15}. In Theorem \ref{theo:composehypo}, we show how to use Lemma \ref{uni} and the techniques used in the proof of Theorem \ref{achievability} and \ref{achievabilityinformed} to reproduce the main result of \cite{BertaBH17} in the case where the set $S_2$ is finite (that is, its size is a constant independent of $n$). 

We also conclude the following as a consequence of Theorem \ref{theo:composehypo} and \cite[Section IV. B]{BertaBH17}. In classical hypothesis testing, it is well known that to distinguish an i.i.d. distribution $P^{\otimes n}$ with another i.i.d. distribution $Q^{\otimes n}$ such that Type $1$ error is $\eps$ and Type $2$ error is $2^{-n\relent{P}{Q} + O(\sqrt{n\log\frac{1}{\eps}})}$, it suffices to take the test (that accepts $P^{\otimes n}$ with high probability) to be the set of typical strings associated to the distribution $P^{\otimes n}$ \cite{Hoeffding65}. Thus, this test is independent of $Q$. We show in Corollary \ref{nouniversaltest} that such a test does not exist in the quantum case.

\section{Preliminaries}
\label{sec:prelims}

For a natural numbers $n,m$ with $n\leq m$, let $[n:m]$ represent the set $\{n,n+1,\ldots m\}$. For $N>0$, $\log N$ is with respect to the base $2$ and $\ln N$ is with respect to base $e$.  

Consider a finite dimensional Hilbert space $\cH$ endowed with an inner product $\langle \cdot, \cdot \rangle$ (in this paper, we only consider finite dimensional Hilbert-spaces). The $\ell_1$ norm of an operator $X$ on $\cH$ is $\| X\|_1:=\Tr\sqrt{X^{\dagger}X}$, $\ell_2$ norm is $\| X\|_2:=\sqrt{\Tr XX^{\dagger}}$ and the $\ell_\infty$ norm is $\|X\|_\infty$, which is the largest eigenvalue of $\sqrt{X^{\dagger}X}$ . Let $\cL(\cH)$ be the set of all linear operators (or matrices) on $\cH$. A quantum state (or a density matrix) is positive semi-definite matrix on $\cH$ with trace equal to $1$. It is called {\em pure} if and only if its rank is $1$. A sub-normalized quantum state is a positive semi-definite matrix on $\cH$ with trace less than or equal to $1$. Let $\ket{\psi}$ be a unit vector on $\cH$, that is $\langle \psi,\psi \rangle=1$.  With some abuse of notation, we use $\psi$ to represent the quantum state and also the density matrix $\ketbra{\psi}$, associated with $\ket{\psi}$. Given a quantum state $\rho$ on $\cH$, {\em support of $\rho$}, called $\text{supp}(\rho)$ is the subspace of $\cH$ spanned by all eigenvectors of $\rho$ with non-zero eigenvalues. The set of all quantum states on a Hilbert space $\cH$ is denoted by $\cD(\cH)$.
 
A {\em quantum register} $A$ is associated with some Hilbert space $\cH_A$. A quantum state $\rho$ on register $A$ is represented as $\rho_A $. If two registers $A,B$ are associated with the same Hilbert space, we shall represent the relation by $A\equiv B$.  Composition of two registers $A$ and $B$, denoted $AB$, is associated with Hilbert space $\cH_A \otimes \cH_B$.  For two quantum states $\rho$ and $\sigma$, $\rho\otimes\sigma $ represents the tensor product (Kronecker product) of $\rho$ and $\sigma$. The identity operator on $\cH_A$ (and associated register $A$) is denoted $\id_A$. For normal operators $P$ and $Q$ we will use the notation $P \preceq Q$ ($P \prec Q$) if $(Q-P)$ is a positive semi-definite operator (positive definite operator). Given a set of quantum states $\{\rho^{(i)}_A\}$ on a register $A$, the set of all convex combinations of quantum states in this set will be represented by $\conv(\{\rho^{(i)}_A\})$.

Let $\rho_{AB}$ be a quantum state. We define $\rho_{B} := \Tr_{A}\rho_{AB}
:= \sum_i (\bra{i} \otimes \id_{B}) \rho_{AB} (\ket{i} \otimes \id_{B}) ,$
where $\{\ket{i}\}_i$ is an orthonormal basis for the Hilbert space $\cH_A$.
The quantum state $\rho_B$ is referred to as the marginal quantum state of $\rho_{AB}$. Unless otherwise stated, a missing register from subscript in a quantum state will represent partial trace over that register. Given a $\rho_A$, a {\em purification} of $\rho_A$ is a pure quantum state $\rho_{AB}$ such that $\Tr_{B}{\rho_{AB}}=\rho_A$. Purification of a quantum state is not unique.

A {\em unitary} operator $U_A:\cH_A \rightarrow \cH_A$ is such that $U_A^{\dagger}U_A = U_A U_A^{\dagger} = \id_A$. An {\em isometry}  $V:\cH_A \rightarrow \cH_B$ is such that $V^{\dagger}V = \id_A$ and $VV^{\dagger} = \id_B$. A POVM on the register $A$ is a collection of positive semi-definite operators $\{M_i\}_{i\in \mathcal{I}}$ such that $\sum_i M_i = \id_A$. A quantum channel $\cN_{A\to B}: \cL(\cH_A)\to \cL(\cH_B)$ is a completely positive and trace preserving map, and it takes a quantum state on $\cD(\cH_A)$ to a quantum state on $\cD(\cH_B)$.

 We shall use the following standard information theoretic quantities. 
\begin{itemize}
\item {\bf Quantum relative entropy:} (\cite{umegaki1954}) Let $\rho,\sigma\in \cD(\cH)$ be quantum states.
$$\relent{\rho}{\sigma}: = \Tr[\rho(\log\rho-\log\sigma)].$$
\item {\bf Quantum relative entropy variance:}  Let $\rho,\sigma\in \cD(\cH)$ be quantum states.
$$\varent{\rho}{\sigma}:=\Tr[\rho(\log\rho-\log\sigma)^2] - \relent{\rho}{\sigma}^2.$$
\item {\bf Quantum mutual information:} Let $\rho_{AB}\in \cD(\cH_{AB})$ be a quantum state. 
$$\mutinf{A}{B}_{\rho_{AB}}=\relent{\rho_{AB}}{\rho_A\otimes \rho_B}.$$
\item {\bf Smooth quantum hypothesis testing divergence:} (\cite{BuscemiD10}, see also \cite{HyashiN03}) Let $\rho,\sigma\in \cD(\cH)$ be quantum states.
$$\dheps{\rho}{\sigma}{\eps} : = \max_{\substack{0\preceq M \preceq \mathbb{I} \\ \tr\left[ M \rho\right] \geq 1-\eps }} -\log \tr \left[M\sigma \right].$$
We refer to $\eps$ as the Type $1$ error and $2^{-\dheps{\rho}{\sigma}{\eps}}$ as the Type $2$ error. 
\item {\bf Max-relative entropy} (\cite{Datta09}) For $\rho,\sigma\in \mathcal{D}(\cH)$ such that $\text{supp}(\rho) \subseteq \text{supp}(\sigma)$, $$ \dmax{\rho}{\sigma}  \defeq  \inf \{ \lambda \in \mathbb{R} :   \rho \preceq 2^{\lambda} \sigma\}  .$$  
\item {\bf Max-information:} (\cite{Renner13}) Let $\rho_{AB}\in \cD(\cH_{AB})$ be a quantum state. $$\imax(A:B)_{\rho_{AB}}:= \dmax{\rho_{AB}}{\rho_A\otimes \rho_B}.$$ 
\end{itemize}

\begin{fact}
\label{dhdmax}
Let $\rho,\sigma,\tau \in \cD(\cH)$ be quantum states and $k>0$ be a real such that $\sigma \preceq 2^k\tau$. Then for any $\eps\in (0,1)$,
$$\dheps{\rho}{\sigma}{\eps} \geq \dheps{\rho}{\tau}{\eps} - k.$$
\end{fact}
\begin{proof}
Let $M$ be the operator that achieves the maximum in the definition of $\dheps{\rho}{\tau}{\eps}$. Then $$2^{-\dheps{\rho}{\sigma}{\eps}}\leq \Tr[M\sigma]\leq 2^k\Tr[M\tau] = 2^{k - \dheps{\rho}{\tau}{\eps}}.$$  
\end{proof}

\begin{fact}
\label{fact:smoothdh}
Let $\eps, \delta\in (0,1)$ and $\rho, \sigma \in \cD(\cH)$ be quantum states such that $\Pur(\rho, \sigma)\leq \delta$. Then for any quantum state $\tau$, 
$$\dheps{\rho}{\tau}{\eps+\delta} \geq \dheps{\sigma}{\tau}{\eps}.$$
\end{fact}
\begin{proof}
Let $\Lambda$ be the operator achieving the supremum in the definition of $\dheps{\sigma}{\tau}{\eps}$. Then $\Tr(\Lambda\sigma)\geq 1-\eps$. This implies that $$\Tr(\Lambda\rho)\geq \Tr(\Lambda\sigma) - \frac{1}{2}\|\rho-\sigma\|_1 \geq 1-\eps- \Pur(\rho, \sigma) \geq 1-\eps-\delta.$$ Further,
$$2^{-\dheps{\sigma}{\tau}{\eps}} = \Tr(\Lambda\sigma) \geq 2^{-\dheps{\rho}{\tau}{\eps+\delta}},$$ by the definition of $\dheps{\rho}{\tau}{\eps+\delta}$. This completes the proof.
\end{proof}

\begin{fact}
\label{marginalbest}
Let $\rho_{AB}\in \cD(\cH_{AB})$ be a quantum state. Then for all quantum states $\tau_A \in \cD(\cH_A),\sigma_B \in \cD(\cH_B)$, $\relent{\rho_{AB}}{\tau_A\otimes\sigma_B} \geq \relent{\rho_{AB}}{\rho_A\otimes\rho_B}$.
\end{fact}
\begin{fact}
\label{varentmaxent}
Let $\rho,\sigma \in \cD(\cH)$ be quantum states such that $\rho \preceq 2^k\sigma$. Then it holds that $\varent{\rho}{\sigma} \leq k^2.$
\end{fact}
\begin{proof}
We notice that 
$$\log\rho-\log\sigma \preceq k\cdot\id+\log\sigma-\log\sigma = k\cdot \id.$$ Since $\log\rho - \log\sigma$ commutes with $\id$, we conclude
$$\Tr[\rho(\log\rho-\log\sigma)^2]  \leq \Tr[\rho(k+\log\sigma-\log\sigma)^2] \leq k^2.$$
\end{proof}

Following quantities are variants of smooth quantum hypothesis testing divergence.

\begin{definition}
\label{ihdefinition}
Let  $\rho_{AB} \in \cD(\cH_{AB})$ be a quantum state. Define: 
\begin{align*}
\Iheps{A}{B}{\eps}_{\rho_{AB}}&:= \min_{\sigma_A} \quad \dheps{\rho_{AB}}{\sigma_A\otimes\rho_B}{\eps}\\& = \min_{\sigma_A} \max_{\substack{0\preceq M \preceq \mathbb{I} \\ \tr\left[ M \rho_{AB}\right] \geq 1-\eps }} - \log \tr \left[ M \left(\sigma_{A} \otimes \rho_B\right)\right].
\end{align*}
\end{definition}

We note that the minimization in above quantity is over the first register in the argument. 

\begin{definition}
\label{ihdefinition2}
Let  $\rho_{AB}\in \cD(\cH_{AB}),\sigma_B \in \cD(\cH_B)$ be quantum states. Let $S_A \subseteq \cD(\cH_A)$ be a convex subset of quantum states on register $A$. Define: 
\begin{align*}
\Ihepstres{A}{B}{\eps}{S_A}_{\rho_{AB},\sigma_B}&:= \min_{\tau_A \in S_A} \dheps{\rho_{AB}}{\tau_A\otimes\sigma_B}{\eps}.\\
\end{align*}
\end{definition}

\begin{definition}
\label{ihdefinition3}
Let  $\rho_{AB}\in \cD(\cH_{AB})$ be a quantum state. Let $S_A \subseteq \cD(\cH_A),S_B \subseteq \cD(\cH_B)$ be two convex subsets of quantum states on registers $A,B$ respectively. Define: 
\begin{align*}
\Ihepshres{A}{B}{\eps}{S_A, S_B}_{\rho_{AB}}&:= \min_{\sigma_B\in S_B} \Ihepstres{A}{B}{\eps}{S_A}_{\rho_{AB},\sigma_B} = \min_{\sigma_B\in S_B,\tau_A\in S_A}\dheps{\rho_{AB}}{\tau_A\otimes\sigma_B}{\eps}.\\
\end{align*}
\end{definition}

Following definition is useful when $S_A,S_B$ are allowed to be set of all quantum states on registers $A$ and $B$ respectively.

\begin{definition}
\label{ihdefinition4}
Let  $\rho_{AB}\in \cD(\cH_{AB})$ be a quantum state.  Define: 
\begin{align*}
\Ihepsh{A}{B}{\eps}_{\rho_{AB}}&:= \min_{\sigma_B,\tau_A}\dheps{\rho_{AB}}{\tau_A\otimes\sigma_B}{\eps}.\\
\end{align*}
\end{definition}

\begin{fact}[Minimax theorem, \cite{vonNeumann1928}]
\label{fact:minimax}
Let $\cX,\cY$ be convex compact sets and $f:\cX\times \cY\rightarrow \mathbb{R}$ be a continuous function that satisfies the following properties: $f(\cdot ,y): \cX\rightarrow \mathbb {R}$ is convex for fixed $y$, and
$f(x,\cdot ):\cY\rightarrow \mathbb {R}$  is concave for fixed $x$. Then it holds that
$$\min_{x\in \cX}\max_{y\in \cY} f(x,y) = \max_{y\in \cY}\min_{x\in \cX} f(x,y).$$
\end{fact}

The following lemma follows from Definition \ref{ihdefinition} and Fact \ref{fact:minimax}. A related result for the trace distance was obtained in \cite{Jain05}.
\begin{lemma}
\label{minimaxoperator}
Let  $\rho_{AB}\in \cD(\cH_{AB})$ be a quantum state. There exists a positive operator $M^*$ satisfying $\Tr[M^*\rho_{AB}] \geq 1-\eps$, such that for all quantum states $\sigma_A \in \cD(\cH_{A})$,
$$\Tr[M^*(\sigma_A\otimes \rho_B)] \leq 2^{-\Iheps{A}{B}{\eps}_{\rho_{AB}}}.$$
\end{lemma}
\begin{proof}
From Definition \ref{ihdefinition}, we conclude that 
\begin{eqnarray*}
2^{-\Iheps{A}{B}{\eps}_{\rho_{AB}}} &=&  \max_{\sigma_A} \min_{\substack{0\preceq M \preceq \mathbb{I} \\ \tr\left[ M \rho_{AB}\right] \geq 1-\eps }}\tr \left[ M \left(\sigma_{A} \otimes \rho_B\right)\right] \\ &\overset{a}=& 
\min_{\substack{0\preceq M \preceq \mathbb{I} \\ \tr\left[ M \rho_{AB}\right] \geq 1-\eps }} \max_{\sigma_A} \tr \left[ M \left(\sigma_{A} \otimes \rho_B\right)\right]\\ &\overset{b} =& \max_{\sigma_A} \tr \left[ M^* \left(\sigma_{A} \otimes \rho_B\right)\right], 
\end{eqnarray*}
where $a$ follows from the minimax theorem (Fact \ref{fact:minimax}) and the facts that $\tr \left[ M \left(\sigma_{A} \otimes \rho_B\right)\right]$ is linear in $M$ for a fixed $\sigma_A$ (and vice-versa), $\sigma_A$ belongs to a convex compact set and $M$ belongs to a convex compact set and $b$ follows by defining $M^*$ to the operator that achieves the infimum in second equality. The lemma concludes with the observation that $M^*$ also satisfies $\Tr[M^*\rho_{AB}]\geq 1-\eps$.

\end{proof}

\begin{fact}[Jordan's lemma, \cite{jordan1875}] \label{Jordan}
For any two projectors $\Pi^{(1)}$ and $\Pi^{(2)}$, there exists a set of orthogonal projectors $\{\Pi_\alpha\}_{\alpha=1}^k$ (each of dimension either one or two), for some natural number $k$, such that 
\begin{enumerate}
\item $\sum_{\alpha\in [1:k]} \Pi_\alpha = \id$.
\item $\Pi_\alpha\Pi^{(i)}= \Pi^{(i)}\Pi_\alpha$, for all $i\in \{1,2\}, \alpha\in [1:k]$. 
\item $\Pi_\alpha\Pi^{(i)}\Pi_\alpha$ is a one dimensional projector, for $i\in \{1,2\}, \alpha \in [1:k]$.
\end{enumerate} 
\end{fact}

\begin{fact}[Hayashi-Nagaoka inequality, \cite{HyashiN03}]
\label{haynag}
Let $c>0$ be a real and $0\prec S \prec \id,T$ be positive semi-definite operators. Then 
$$\id- (S+T)^{-\frac{1}{2}}S(S+T)^{-\frac{1}{2}}\preceq (1+c)(\id-S) + (2+c+\frac{1}{c})T.$$
\end{fact}

\begin{fact} [Neumark's theorem, \cite{Watrouslecturenote}] \label{Neumark}For any POVM $\left\{M_i\right\}_{i \in \mathcal{I}}$ acting on a system $S,$ there exists a unitary $U_{SP}$ and an orthonormal basis $\left\{\ket{i}_P\right\}_{i \in \mathcal{I}}$ such that for all quantum states $\rho_S$, we have
$$\tr \left[U^\dagger_{SP} \left(\mathbb{I}_S \otimes \ket{i}\bra{i}_P\right)U_{SP}\left(\rho_S \otimes \ket{0}\bra{0}_P\right)\right] = \tr \left[ M_i \rho_S\right].$$
\end{fact}

\section{A union of projectors}
\label{sec:quantumunion}
In this section, we prove a quantum version of the following classical statement. Let $E_1, E_2 \subseteq \cX$ be two sets and $p$ be a probability distribution over $\cX$. Then there exists a set $E^* \subseteq \cX$ (more precisely, $E^* = E_1\cup E_2$) such that $\Pr_p\left\{E^*\right\} \geq \max\{\Pr_p\left\{E_1\right\},\Pr_p\left\{E_2\right\}\}$. The following lemma is a quantum version of this statement.  
\begin{lemma}
\label{lem:jordanunion}
Let $\{\Pi^{(1)}, \Pi^{(2)}\}$ be two projectors. For every $\delta>0$, there exists a projector $\Pi^\star$ such that for all quantum states $\rho \in \cD(\cH)$,
\begin{align}
\label{p11}
\tr \left[\Pi^{\star} \rho \right]& \geq \max\{\Tr[\Pi^{(1)}\rho], \Tr[\Pi^{(2)}\rho]\} - \delta; \\
\label{p22}
\Pi^{\star} & \preceq \frac{2} {\delta^2}\left( \Pi^{(1)} +  \Pi^{(2)} \right) .
\end{align}
\end{lemma}
\begin{proof}
Let $\left\{\Pi_\alpha\right\}_{\alpha=1}^k$ be the set of orthogonal projectors (each either one or two dimensional) obtained by Jordan's lemma (Fact \ref{Jordan}) applied on $\Pi^{(1)},\Pi^{(2)}$ such that $\sum_{\alpha =1}^k \Pi_{\alpha} = \mathbb{I}.$ Furthermore, for $\alpha \in [1:k]$ and $i \in \left\{1,2\right\}$ let 
\begin{align*}
\Pi^{(i)}(\alpha) &:= \Pi_\alpha \Pi^{(i)} \Pi_{\alpha}.
\end{align*}
Observe that $\Pi^{(i)} = \sum_{\alpha} \Pi^{(i)}(\alpha)$. Also, let $$\mathsf{Far}:= \left\{\alpha: \tr\left[\Pi^{(1)}(\alpha) \Pi^{(2)}(\alpha)\right] <  1- \delta^2\right\},$$ and let the set $\mathsf{Near}$ be the compliment  of the set $\mathsf{Far}$.
For every $\alpha \in [1:k],$ let $\Pi^\star(\alpha)$ be defined as follows:
\beq
\label{pistar}
\Pi^\star(\alpha)=
\begin{cases}
\Pi_\alpha & \mbox{if} ~ \alpha \in \mathsf{Far}; \\
 \Pi^{(1)}(\alpha)         & \mbox{otherwise}.
\end{cases}
\enq
We now show that $\Pi^\star:= \sum_{\alpha \in [1:k]}\Pi^\star(\alpha)$ satisfies the properties mentioned in Equations \eqref{p11} and \eqref{p22}. 

\vspace{0.1in}

\noindent {\bf Proof of Equation \eqref{p11}}: Fix a quantum state $\rho$ and let $\rho(\alpha) := \Pi_\alpha\rho\Pi_\alpha$. Let $i\in \{1,2\}$ be such that $\Tr\left[\Pi^{(i)}\rho\right] = \max\{\Tr\left[\Pi^{(1)}\rho\right], \Tr\left[\Pi^{(2)}\rho\right]\}$. Consider,
\begin{align*}
\tr\left[\Pi^\star \rho\right] &= \sum_{\alpha \in \mathsf{Far}}\tr\left[\Pi^\star(\alpha) \rho(\alpha)\right] + \sum_{\alpha \in \mathsf{Near}}\tr\left[\Pi^\star(\alpha) \rho(\alpha)\right]\\
&\overset{a}= \sum_{\alpha \in \mathsf{Far}}\tr\left[\Pi_\alpha \rho(\alpha)\right] + \sum_{\alpha \in \mathsf{Near}}\tr\left[\Pi^{(1)}(\alpha) \rho(\alpha)\right]\\
&\overset{b}\geq \sum_{\alpha \in \mathsf{Far}}\tr\left[\Pi^{(i)}(\alpha) \rho(\alpha)\right] + \sum_{\alpha \in \mathsf{Near}}\tr\left[\Pi^{(1)}(\alpha) \rho(\alpha)\right]\\
& = \sum_{\alpha \in [1:k]}\tr\left[\Pi^{(i)}(\alpha) \rho(\alpha)\right] + \sum_{\alpha \in \mathsf{Near}}{\tr [\rho(\alpha)]}\tr\left[\left(\Pi^{(1)}(\alpha) - \Pi^{(i)}(\alpha) \right)\frac{\rho(\alpha)}{{\tr [\rho(\alpha)]}}\right] \\
& \overset{c} \geq \Tr\left[\Pi^{(i)}\rho\right] -\sum_{\alpha \in \mathsf{Near}}{\tr [\rho(\alpha)]}\left\|\Pi^{(1)}(\alpha) - \Pi^{(i)}(\alpha)\right\|_\infty\\
& \overset{d} \geq \Tr\left[\Pi^{(i)}\rho\right] - \delta\sum_{\alpha \in \mathsf{Near}}{\tr [\rho(\alpha)]} \\
& \geq \Tr[\Pi^{(i)}\rho] - \delta,
\end{align*} 
where $a$ follows from the definition of $\Pi^\star({\alpha})$ mentioned in Equation \eqref{pistar}; $b$ follows from the identity $\Pi_\alpha\succeq \Pi^{(i)}(\alpha)$; $c$ follows since $\sum_{\alpha \in [1:k]}\tr\left[\Pi^{(i)}(\alpha) \rho(\alpha)\right] = \tr\left[\Pi^{(i)} \rho\right]$ and $d$ follows from the property of the set $\mathsf{Near}$.

\vspace{0.1in}

\noindent {\bf Proof of Equation \eqref{p22}:} To prove the property mentioned in Equation \eqref{p22} we assume the following claim:
\begin{claim}
\label{uni1}
For $\alpha \in [1:k]$, $$\Pi^\star(\alpha) \preceq \frac{2}{\delta^2} \left( \Pi^{(1)}(\alpha) +  \Pi^{(2)}(\alpha) \right).$$
\end{claim}
The proof of this claim is given towards the end of this  proof. Notice the following set of inequalities:  
\begin{align*}
\Pi^{\star} &= \sum_{\alpha} \Pi^{\star}(\alpha) \\
 & \preceq \sum_{\alpha}\frac{2}{\delta^2} \left( \Pi^{(1)}(\alpha) +  \Pi^{(2)}(\alpha) \right) \\
& = \frac{2} {\delta^2}  \left( \Pi^{(1)} +  \Pi^{(2)} \right) .
\end{align*}
This completes the proof of the lemma.

\vspace{0.1in}

\noindent {\bf Proof of Claim \ref{uni1}}: Claim trivially follows for $\alpha \in$ $\mathsf{Near}$. We now consider the case when $\alpha \in$ $\mathsf{Far}$. Towards this notice that $\Pi^i(\alpha)$ is a one dimensional projector as guaranteed by Jordan's lemma (Fact \ref{Jordan}), for each $i\in \{1,2\}$. Further, let $\Pi^{(2)}(\alpha)$ be defined as follows:
\beq
\ket{\Pi^{(2)}(\alpha)} = \gamma \ket{\Pi^{(1)}(\alpha)} + \beta \ket{\Pi^1(\alpha)}_{\perp};
\enq
where $\ket{\Pi^{(1)}(\alpha)}_{\perp}$ is the unit vector orthogonal to $\ket{\Pi^{(1)}(\alpha)}$ in the subspace corresponding to $\Pi_{\alpha}$. From the definition of the set $\mathsf{Far}$ we conclude that $|\gamma| < \sqrt{1- \delta^2}.$ Now consider the operator $\Pi^{(1)}(\alpha) +  \Pi^{(2)}(\alpha)$ which can be represented as follows:
\beq
\label{matrix}
\begin{bmatrix}
   1+|\gamma|^2       & \gamma^\star\beta  \\
   \gamma\beta^\star & 1- |\gamma|^2
\end{bmatrix},
\enq
where in the above we have used the fact that $|\beta|^2 = 1- |\gamma|^2.$ The characteristic equation of the matrix in Equation \eqref{matrix} satisfies the following: 
\beq
\label{characteristic}
(1-\lambda)^2 - |\gamma|^4 - |\gamma|^2 (1-|\gamma|^2 ) = 0,
\enq 
where $\lambda$ is an eigenvalue. From Equation \eqref{characteristic} we have that $\lambda \geq 1- |\gamma| > \frac{\delta^2}{2}.$ Thus, $\frac{\delta^2}{2}\Pi_\alpha\preceq\Pi^{(1)}(\alpha) +  \Pi^{(2)}(\alpha)$. This proves the claim.

\end{proof}

Lemma \ref{lem:jordanunion} allows us to the prove the following lemma, which we shall use in our main results.
\begin{lemma}
\label{uni}
Let $\eps,\delta >0$. For $i \in [1:s],$ let $\rho^{(i)}\in \cD(\cH)$ be a quantum state and $\Pi^{(i)}$ be a projection operator such that 
$$\tr \left[\Pi^{(i)} \rho^{(i)} \right] \geq 1- \eps.$$
Then there exists a projection operator $\Pi^\star$ such that 
\begin{align*}
\tr \left[\Pi^{\star} \rho^{(i)} \right]& \geq 1- \eps- (\delta \log(2s)); \\
\Pi^{\star} & \preceq \left(\frac{2} {\delta^2}\right)^{\log(2s)}\left( \Pi^{(1)} +  \Pi^{(2)} + \ldots+ \Pi^{(s)}\right) .
\end{align*}
\end{lemma}
\begin{proof}
Without loss of generality, we assume that $s = 2^{t}$ for some integer $t$. The proof for general $s$ proceeds in similar fashion. We group the projectors $\Pi^{(i)}$ into pairs $\{\Pi^{(1)},\Pi^{(2)}\}, \{\Pi^{(3)},\Pi^{(4)}\},\ldots ,\{\Pi^{(s-1)},\Pi^{(s)}\}$. Applying Lemma \ref{lem:jordanunion} to each of the pair with given $\delta$, we obtain a collection of projectors $\{\Pi^{(1,2)}, \Pi^{(3,4)}, \ldots \Pi^{(s-1,s)}\}$ such that for every odd $i < s$, we have 
$$\Tr \left[\Pi^{(i,i+1)} \rho^{(i)} \right] \geq 1-\eps-\delta,\quad \Tr \left[\Pi^{(i,i+1)} \rho^{(i+1)} \right] \geq 1-\eps-\delta, \quad\Pi^{(i,i+1)} \preceq  \left(\frac{2}{\delta^2}\right)\left( \Pi^{(i)} +  \Pi^{(i+1)}\right) .$$
Now, we further group the projectors $\{\Pi^{(1,2)}, \Pi^{(3,4)}, \ldots \Pi^{(s-1,s)}\}$ into consecutive pairs and by applying Lemma \ref{lem:jordanunion} with given $\delta$, obtain projectors $\{\Pi^{(1,2,3,4)}, \Pi^{(5,6,7,8)}, \ldots \Pi^{(s-3,s-2,s-1,s)}\}$ such that for every $i$ satisfying $i \text{ mod }4=1$, we have
$$\Tr \left[\Pi^{(i,i+1,i+2,i+3)} \rho^{(j)} \right] \geq 1-\eps-2\delta \quad \forall j\in [i,i+3];$$ $$\Pi^{(i,i+1,i+2,i+3)} \preceq  \left(\frac{2} {\delta^2}\right)\left( \Pi^{(i,i+1)} +  \Pi^{(i+2,i+3)}\right) \preceq \left(\frac{2} {\delta^2}\right)^2\left( \Pi^{(i)} +  \Pi^{(i+1)} + \Pi^{(i+2)} + \Pi^{(i+3)}\right).$$
Continuing in this way till $\log(s)$ steps (for a general $s$, the number of steps will be at most $\log (2s)$), we obtain the desired projector $\Pi^*$. This completes the proof.
\end{proof}

\section{Communication over compound quantum channel}
\label{sec:channel}
\begin{definition}
\label{code}
Let $\ket{\theta}_{E_AE_B}$ be the shared entanglement between Alice  and Bob. Let $M$ be the message register. An $(R, \eps )$-entanglement assisted code for a compound quantum channel $\left\{\cN^{(i)}_{ A \to B}\right\}_{i=1}^s$ consists of 
\begin{itemize}
\item An encoding operation $\mathcal{E}: ME_A \rightarrow A $ for Alice .  
\item A decoding operation $\mathcal{D} : B E_B\rightarrow M'$ for Bob, with $M'\equiv M$ being the output register such that for all $m$ and for all $i \in [1:s]$,
\beq
\Pr\left\{M'\neq m|M=m\right\} \leq \eps. \nonumber
\enq
\end{itemize}
\end{definition}

\subsection*{Achievability result}

\begin{theorem}
\label{achievability}
Let $\left\{\cN^{(i)}_{A \to B }\right\}_{i=1}^s$ be a compound quantum channel and let $\eps,\eta \in (0,1)$. Let $A'\equiv A$ be a purifying register. Then, for any $R$ satisfying 
\beq
\label{pointchannelrate}
R \leq\max _{\ketbra{\psi}_{A A'}} \left( \min_{i \in [1:s]} \Iheps{B}{A'}{\eps}_{\cN^{(i)}_{A \to B } (\ketbra{\psi}_{AA'})} + (2\log2s)\log \left(\frac{\eta}{6\log(2s)}\right) + \log\frac{\eps}{4s}\right),
\enq
there exists an $(R,\eps+3\eta)$-entanglement assisted code for the compound quantum channel $\left\{\cN^{(i)}_{A \to B }\right\}_{i=1}^s.$
\end{theorem}

\vspace{0.1in}

\noindent \emph{Outline of the protocol:} In the protocol, Alice and Bob share many copies of a quantum state $\ketbra{\psi}_{AA'}$, one copy corresponding to each message. To send a message, Alice communicates the corresponding share of her state through the channel. Bob performs the position-based decoding strategy to obtain the correct message with high probability. Since Bob does not know which channel has been used, he uses Lemma \ref{uni} to obtain a single operator that gives small error (of Type $1$) in the quantum hypothesis testing, for all channels. 

\begin{proof}[Proof of Theorem \ref{achievability}]
Fix $\ketbra{\psi}_{AA'}$ and $R$ as given in Equation \eqref{pointchannelrate}. Introduce the registers $A_1,A_2,\ldots A_{2^R}$, such that $A_i\equiv A$ and $A'_1,A'_2,\ldots A'_{2^R}$ such that $A'_i\equiv A'$.  Alice  and Bob  share the quantum state $$\ketbra{\psi}_{A_1A'_1}\otimes \ketbra{\psi}_{A_2A'_2},\ldots \ketbra{\psi}_{A_{2^R}A'_{2^R}},$$
 where Alice  holds the registers $A_1,A_2, \cdots, A_{2^{R}}$ and Bob  holds the registers $A'_1,A'_2, \cdots, A'_{2^{R}}$. For $i \in [1:s],$ let $ 0 \preceq M^{(i)}_{BA'} \preceq \mathbb{I}$ be such that for all $j \in [1:s]$, we have
\beq
\label{optimalmeasurement}
\Iheps{B}{A'}{\eps}_{\cN^{(i)}_{A \to B } (\ketbra{\psi}_{AA'})}\leq - \log \tr \left[M^{(i)}_{BA'}\left( \cN^{(j)}_{A\to B}(\psi_{A}) \otimes \psi_{A'}\right) \right ]. \nonumber
\enq
The existence of such an operator $M^{(i)}_{BA'}$ is guaranteed by Lemma \ref{minimaxoperator}. Further, as guaranteed by the Neumark's  theorem (Fact \ref{Neumark}), let $\Pi^{(i)}_{BA'P}$ be such that $\forall i, j \in [1:s],$ $$ \tr \left[M^{(i)}_{BA'} \left( \cN^{(j)}_{A\to B}(\psi_{A}) \otimes \psi_{A'}\right) \right ] = \tr \left[\Pi^{(i)}_{BA'P} \left( \cN^{(j)}_{A\to B}(\psi_{A}) \otimes \psi_{A'} \otimes \ketbra{0}_P\right) \right ] ,$$ and  $$ \tr \left[M^{(i)}_{BA'} \left(\cN^{(i)}_{A\to B} (\ketbra{\psi}_{AA'})\right) \right ] = \tr \left[\Pi^{(i)}_{BA'P} \left(\cN^{(i)}_{A\to B} (\ketbra{\psi}_{AA'}) \otimes \ketbra{0}_P\right) \right ]. $$ Let $\Pi^\star_{BA'P}$  be the operator obtained by setting $$\frac{\eta}{3\log(2s)}\leftarrow \delta, \quad \cN^{(i)}_{A\to B}(\ketbra{\psi}_{AA'}) \otimes\ketbra{0}_P \leftarrow \rho^{(i)}, \quad\Pi^{(i)}_{BA'P} \leftarrow \Pi^{(i)}$$ in Lemma \ref{uni}.

Our protocol is as follows:
\vspace{2mm}

\noindent {\bf{Encoding:}} Alice  on receiving the message $m \in [1:2^R]$ sends the register $A_m$ over the channel. Assuming that the channel $\cN^{(i)}_{A\to B}$ was used for transmission, the quantum state in Bob's possession is the following:
\begin{equation*}
{{\Theta}}^{(i)}_{ BA'_1\cdots A'_{2^{R}}}:= \psi_{A'_1}\otimes  \cdots\otimes \cN^{(i)}_{ A_m \to B}\left( \ketbra{\psi}_{A_mA'_m}\right)\cdots \otimes \psi_{A'_{2^R}},
\end{equation*}
for some $i \in [1:s].$ Further, notice that the quantum state ${\Theta}^{(i)}_{BA^\prime_j}$ between the register $A'_j$ and the channel output $B$ is the following  
\begin{equation*}
\label{jointstate}
{{\Theta}}^{(i)}_{BA^\prime_j}   =
\begin{cases}
\cN^{(i)}_{A \to B} (\ketbra{\psi}_{AA'}) & \mbox{if }~ j =m; \\
\cN_{A \to B}^{(i)}(\psi_{A}) \otimes \psi_{A'}        & \mbox{otherwise.}
\end{cases}
\end{equation*}
 {\bf{Decoding:}} The decoding technique is derived from the work \cite{AnshuJW17}. For each $m \in [1:2^R]$ consider the following operator
\begin{equation}
\Lambda(m):= \mathbb{I}_{A'_1} \otimes \mathbb{I}_{A'_2} \otimes \cdots \Pi^\star_{BA'_mP} \otimes \cdots \otimes \mathbb{I}_{A'_{2^R}}, \nonumber
\end{equation}
where $\Pi^\star_{BA'_mP}$ is as discussed above. The decoding POVM element corresponding to $m $ is:
\begin{equation*}
\Omega(m) := \left(\sum_{ m^\prime \in [1:2^{R}]} \Lambda({m'})\right)^{-\frac{1}{2}}\Lambda({m})\left(\sum_{ m^\prime \in [1:2^{R}]} \Lambda({m'})\right)^{-\frac{1}{2}}. 
\end{equation*}
It is easy to observe that $\sum_m \Omega(m) \preceq \mathbb{I}$. Bob on receiving the channel output appends an ancilla $\ket{0}\bra{0}_P$ to his registers and then measures his registers using the POVM defined above. He outputs `$0$' for the outcome corresponding to the POVM element $\id - \sum_m \Omega(m)$.

\vspace{2mm}

\noindent {\bf{Probability of error:}} Let $M$ be the message which was transmitted by Alice  using the strategy above and let $M'$ be the decoded message by Bob  using the above mentioned decoding POVM. Further, let us assume that the channel $\cN^{(i)}_{ A \to B}$ be used for this transmission. By the symmetry of the encoding and decoding strategy, it is enough to show that $\Pr \left\{M' \neq 1 \mid  M=1 \right\} \leq \eps+3\eta$, under the event that $M=1$ is the transmitted message. 
\begin{align*}
& \Pr \left\{M' \neq 1 | M=1\right\} \\ & = \tr \left[\left(\mathbb{I} -\Omega(1)\right){\Theta}^{(i)}_{B A^\prime_1A^\prime_1\cdots A^\prime_{2^{R}}} \otimes \ketbra{0}_P\right]\\
& \overset{a} \leq   \frac{\eps+2\eta}{\eps+\eta}\cdot\tr \left[\left(\mathbb{I} -\Lambda(1)\right){\Theta}^{(i)}_{B A^\prime_1A^\prime_1\cdots A^\prime_{2^{R}}}\otimes \ketbra{0}_P\right] + (4+\frac{\eta}{\eps})\sum_{m'\neq 1} \Tr \left[\Lambda(m'){\Theta}^{(i)}_{B A^\prime_1A^\prime_1\cdots A^\prime_{2^{R}}} \otimes \ketbra{0}_P \right]\\
& \overset{b} =   \frac{\eps+2\eta}{\eps+\eta}\cdot\tr \left[\left(\mathbb{I} - \Pi^\star_{BA'_1P}\right){\Theta}^{(i)}_{B A^\prime_1}\otimes \ketbra{0}_P\right] + (4+\frac{\eta}{\eps})\sum_{m'\neq 1} \Tr \left[\Pi^\star_{BA'_{m'}P}{\Theta}^{(i)}_{B A^\prime _{m'}}\otimes \ketbra{0}_P \right]\\
&\overset{c} \leq \frac{\eps+2\eta}{\eps+\eta}(\eps + \eta) + \frac{4\eta}{\eps}s\times \left(\frac{6\log(2s)} {\eta}\right)^{2\log(2s)} \times  \text{exp}\left(\ln 2\cdot\left(R -  \min_{j\in [1:s]}\Iheps{B}{A'}{\eps}_{\cN^{(j)}_{A \to B } (\ketbra{\psi}_{AA'})}\right)\right)\\
& \overset{d}\leq \eps + 3\eta .
\end{align*}
where $a$ follows from Hayashi-Nagaoka operator inequality (Fact \ref{haynag}) by choosing $c=\frac{\eta}{\eps+\eta}$; $b$ follows from the definition of $\Lambda (m)$, $c$ follows from the properties of $\Pi^\star_{BA'P}$ (see Lemma \ref{uni}) and $d$ follows from our choice of $R.$
\end{proof}

\noindent \textbf{Necessity of the $\log s$ terms:} We give an example of a compound quantum channel where it is necessary to have the $\log s$ terms in Theorem \ref{achievability}. Let $A \equiv B$ and $s = |A|^4$. Decompose the register $A$ into $\log|A|$ qubits. Let $\{\cN^{(i)}_{A\to A}\}_{i=1}^{|A|^4} = \{\id, X, Y, Z\}^{\otimes \log|A|}$, that is, each channel $\cN^{(i)}_{A\to A}$ is a product of Pauli operators acting on the qubits (and hence a unitary operator). Letting $\ketbra{\psi}_{AA'}$ to be the maximally entangled state, we obtain      
$$\min_{i \in [1:s]} \Iheps{B}{A'}{\eps}_{\cN^{(i)}_{A \to B } (\ketbra{\psi}_{AA'})} = \Iheps{B}{A'}{\eps}_{\ketbra{\psi}_{AA'}} = 2\log|A| + \log(1-\eps) = \frac{1}{2}\log s + \log(1-\eps).$$
On the other hand, a protocol that achieves a reliable communication over all of the channels $\{\cN^{(i)}_{A\to A}\}_{i=1}^{|A|^4}$ also achieves a reliable communication over the average channel $\frac{1}{|A|^4}\sum_{i=1}^{|A|^4}\cN^{(i)}_{A\to A}$. This channel takes any input to the maximally mixed state, and hence has no capacity. As a result, the capacity of this compound quantum channel is equal to zero. This shows that it is not possible to obtain an improved version of Theorem \ref{achievability} without the $-\log s$ terms.

\suppress{

\vspace{0.1in}

\noindent \textbf{Comparison with the one-shot bounds in \cite{BertaGW17}:} The achievability bound for the compound quantum channel in \cite{BertaGW17} was in terms of the smooth min-entropy and smooth conditional max-entropy. We show the following claim, which gives a direct comparison with the achievability bound obtained in Theorem \ref{achievability}. 
\begin{claim}
Let $\rho_{AB}\in \cD(\cH_{AB})$ be a quantum state and $\eps, \delta, \eta, \gamma \in (0,1)$. It holds that
$$\hmineps{A}{\eta}_{\rho} - \hmaxeps{A}{B}{\gamma}_{\rho} \leq$$
\end{claim}
\begin{proof}
Let $\rho'_A$ be the quantum state achieving the optimum in the definition of $\hmineps{A}{\eta}_{\rho}$. It holds that $\rho'_A \preceq (1+\eta)\rho_A$, as $\rho'_A$ is obtained from $\rho_A$ by removing some eigenvalues of the latter. Let $\rho''_{AB}, \sigma_B$ be the quantum states achieving the optimum in the definition of $\hmaxeps{A}{B}{\gamma}_{\rho}$. We have
$$\log F^2(\rho''_{AB}, \rho'_A \otimes \sigma_B)$$
\end{proof}
}

\subsection*{Converse bound}

We introduce the following definition for our converse bound.
\begin{definition}
\label{ptopcode}
Let $\ket{\theta}_{E_AE_B}$ be the shared entanglement between Alice  and Bob. Let $M$ be the message register with uniform distribution over the messages. An $(R, \eps, \psi_A )$-entanglement assisted code for a quantum channel $\cN_{ A \to B}$ consists of 
\begin{itemize}
\item An encoding operation $\mathcal{E}: ME_A \rightarrow A $ for Alice, such that the quantum state at the input register $A$, averaged over the message in register $M$, is $\psi_A$.  
\item A decoding operation $\mathcal{D} : B E_B\rightarrow M'$ for Bob, with $M'\equiv M$ being the output register such that for all $i \in [1:s]$,
\beq
\Pr\left\{M'\neq M\right\} \leq \eps. \nonumber
\enq
\end{itemize}
\end{definition}

\begin{theorem}
\label{theo:converse}
Let $\left\{\cN^{(i)}_{A \to B }\right\}_{i=1}^s$ be a compound quantum channel and let $\eps \in (0,1)$. For any $(R,\eps)$-entanglement assisted code for this compound quantum channel, it holds that 
$$R \leq \max _{\ketbra{\psi}_{A A'}} \left( \min_{i \in [1:s]} \Iheps{B}{A'}{\eps}_{\cN^{(i)}_{A \to B } (\ketbra{\psi}_{AA'})} \right).$$
\end{theorem}
\begin{proof}
We consider the case where the message of Alice is drawn from a uniform distribution. This implies the result in the theorem, which is for the worst case over the messages. Fix an $i \in [1:s]$ and the corresponding quantum channel $\cN^{(i)}_{A\to B}$. As shown in \cite[Equation 76]{MatthewsW14}, for any $(R, \eps, \psi_A)$-entanglement assisted code for the quantum channel $\cN^{(i)}_{A\to B}$, we have $$R\leq \Iheps{B}{A'}{\eps}_{\cN^{(i)}_{A \to B } (\ketbra{\psi}_{AA'})},$$ where $\ketbra{\psi}_{AA'}$ is a purification of $\psi_A$. In the case of compound quantum channel, the value of $i$ is unknown to the communicating parties. Hence, for any $(R,\eps)$- entanglement assisted code for the compound channel $\left\{\cN^{(i)}_{A \to B }\right\}_{i=1}^s$ with average quantum state $\psi_A$ at the input, we have $$R\leq \min_{i\in [1:s]}\Iheps{B}{A'}{\eps}_{\cN^{(i)}_{A \to B } (\ketbra{\psi}_{AA'})}.$$ Since the choice of the quantum state $\psi_A$ is arbitrary, maximizing above expression over all possible pure quantum states $\ketbra{\psi}_{AA'}$ establishes the result.
\end{proof}

Finally, we relate $\Iheps{B}{A'}{\eps}_{\rho_{BA'}}$ to the quantity $\dheps{\rho_{BA'}}{\rho_B\otimes\rho_{A'}}{\eps}$ in the following lemma. 
\begin{lemma}
\label{theo:Ihdhsame}
Let $\rho_{BA'}$ be a quantum state and $\eps \in (0,1)$. For every $\delta>0$, it holds that 
$$\dheps{\rho_{BA'}}{\rho_B\otimes \rho_{A'}}{\eps}- 2\log\frac{\eps}{\delta} \leq \Iheps{B}{A'}{\eps+\delta}_{\rho_{BA'}} \leq \dheps{\rho_{BA'}}{\rho_B\otimes \rho_{A'}}{\eps + \delta}.$$
\end{lemma}
\begin{proof}
Consider a purification $\ketbra{\rho}_{ABA'}$ of the quantum state $\rho_{BA'}$. Let $\cN_{AB\to B}$ be the channel that traces out register $A$. As discussed in \cite[Theorem 7]{QiWW17}, the protocol in \cite[Theorem 1]{AnshuJW17} shows that there exists an $(R, \eps+\delta, \rho_{AB})$-entanglement assisted code for the channel $\cN_{AB\to B}$ such that
$$R = \dheps{\cN_{AB\to B}(\rho_{ABA'})}{\cN_{AB\to B}(\rho_{AB})\otimes \rho_{A'}}{\eps} - 2\log\frac{\eps}{\delta} = \dheps{\rho_{BA'}}{\rho_B\otimes \rho_{A'}}{\eps} - 2\log\frac{\eps}{\delta}.$$
On the other hand, as shown in \cite[Equation 76]{MatthewsW14}, for any $(R,\eps+\delta, \rho_{AB})$-entanglement assisted code for the channel $\cN_{AB\to B}$, it holds that 
$$R \leq \Iheps{B}{A'}{\eps+\delta}_{\cN_{AB\to B}(\rho_{ABA'})} = \Iheps{B}{A'}{\eps+\delta}_{\rho_{BA'}}.$$ This establishes the lower bound on $\Iheps{B}{A'}{\eps+\delta}_{\rho_{BA'}}$. The upper bound on $\Iheps{B}{A'}{\eps+\delta}_{\rho_{BA'}}$ follows from the definition. This completes the proof.
\end{proof}

\subsection*{Asymptotic and i.i.d. analysis}

We now show the asymptotic and i.i.d. behavior of our achievability result, reproducing the bound obtained in \cite{BertaGW17}.
\begin{corollary}
\label{cor:asympcomp}
Let $\left\{\cN^{(i)}_{A \to B }\right\}_{i=1}^s$ be a compound quantum channel. Let $A'\equiv A$ be a purifying register. Let $C$ satisfy 
\beq
C \leq\max _{\ketbra{\psi}_{A A'}} \left( \min_{i \in [1:s]} \mutinf{B}{A'}_{\cN^{(i)}_{A\to B}(\Psi_{AA'})}\right).
\enq
For every $\eps \in (0,1), \delta>0$, there exists a large enough $n$ such that there exists a $(n(C - \delta), 2\eps)$-entanglement assisted code for the compound quantum channel $\left\{\cN^{(i) \otimes n}_{A \to B }\right\}_{i=1}^s.$
\end{corollary}
\begin{proof}
From Theorem \ref{achievability}, there exists a $(R, 2\eps)$-entanglement assisted code for the compound quantum channel $\left\{\cN^{(i) \otimes n}_{A \to B }\right\}_{i=1}^s$ for any $R$ satisfying  
$$R \leq\max _{\ketbra{\psi}_{A A'}} \left( \min_{i \in [1:s]} \Iheps{B^n}{A'^n}{\eps}_{\cN^{(i) \otimes n}_{A \to B } (\ketbra{\psi}^{\otimes n}_{AA'})} + (2\log2s)\log \left(\frac{\eta}{6\log(2s)}\right) + \log\frac{\eps}{4s}\right),$$
where $A^n$ and $B^n$ are $n$ copies of the register $A, B$ respectively. From Lemma \ref{theo:Ihdhsame}, it suffices to have 
$$R \leq\max _{\ketbra{\psi}_{A A'}} \left( \min_{i \in [1:s]} \dheps{\cN^{(i) \otimes n}_{A \to B}(\ketbra{\psi}^{\otimes n}_{AA'})}{\cN^{(i) \otimes n}_{A \to B}(\ketbra{\psi}^{\otimes n}_{A})\otimes \psi_{A'}^{\otimes n}}{\eps} - (2\log \frac{2s}{\eps})\log \left(\frac{18\log(2s)}{\eps}\right)\right).$$
Using the asymptotic and i.i.d. behavior of the smooth quantum hypothesis testing divergence given in \cite[Equation34]{TomHay13} and \cite{li2014}, we conclude that it suffices to have $R$ smaller than
$$\max _{\ketbra{\psi}_{A A'}} \left( \min_{i \in [1:s]} n\relent{\cN^{(i)}_{A \to B}(\ketbra{\psi}_{AA'})}{\cN^{(i)}_{A \to B}(\ketbra{\psi}_{A})\otimes \psi_{A'}} - O\left(\sqrt{n\log\frac{1}{\eps}}\right) - (2\log \frac{2s}{\eps})\log \left(\frac{18\log(2s)}{\eps}\right)\right).$$
Choosing $n$ large enough such that $\delta \geq O(\sqrt{\frac{1}{n}\log\frac{1}{\eps}}) + \frac{2\log \frac{2s}{\eps}}{n}\log \left(\frac{18\log(2s)}{\eps}\right)$, the proof concludes. 
\end{proof}

\section{The case of informed sender}
\label{sec:informedsender}
In this section we discuss the case where the sender is aware about which channel in the set $\left\{\cN^{(i)}_{A \to B}\right\}_{i=1}^s$ is being used for transmission. The following lemma is an analogue of Lemma \ref{minimaxoperator} and follows from Definition \ref{ihdefinition2} and Fact \ref{fact:minimax}.
\begin{lemma}
\label{minimaxoperator2}
Let  $\rho_{AB},\sigma_B$ be quantum states and $S_A$ be a convex set of quantum states on register $A$. There exists an operator $M^*$ satisfying $\Tr[M^*\rho_{AB}] \geq 1-\eps$, such that for all quantum states $\tau_A\in S_A$,
$$\Tr[M^*(\tau_A\otimes \sigma_B)] \leq 2^{-\Ihepstres{A}{B}{\eps}{S_A}_{\rho_{AB},\sigma_B}}.$$
\end{lemma}
\begin{proof}
From Definition \ref{ihdefinition2}, we conclude that 
\begin{eqnarray*}
2^{-\Ihepstres{A}{B}{\eps}{S_A}_{\rho_{AB},\sigma_B}} &=&  \max_{\tau_A\in S_A} \min_{\substack{0\preceq M \preceq \mathbb{I} \\ \tr\left[ M \rho_{AB}\right] \geq 1-\eps }}\tr \left[ M \left(\tau_{A} \otimes \sigma_B\right)\right] \\ &\overset{a}=& 
\min_{\substack{0\preceq M \preceq \mathbb{I} \\ \tr\left[ M \rho_{AB}\right] \geq 1-\eps }} \max_{\tau_A \in S_A} \tr \left[ M \left(\tau_{A} \otimes \sigma_B\right)\right]\\ &\overset{b} =& \max_{\tau_A\in S_A} \tr \left[ M^* \left(\tau_{A} \otimes \sigma_B\right)\right],
\end{eqnarray*}
where $a$ follows from the minimax theorem (Fact \ref{fact:minimax}) and the facts that $\tr \left[ M \left(\tau_{A} \otimes \sigma_B\right)\right]$ is linear in $M$ for a fixed $\tau_A$ (and vice versa), $\tau_A$ belongs to the convex compact set $S_A$ and $M$ belongs to a convex compact set and $b$ follows by defining $M^*$ to the operator that achieves the infimum in second equality. The lemma concludes with the observation that $\Tr[M^*\rho_{AB}]\geq 1-\eps$.

\end{proof}

\begin{definition}
\label{informedsendercode}
Let $\ket{\theta}_{E_AE_B}$ be the shared entanglement between Alice  and Bob. Let $M$ denote the message register. An $(R, \eps )$-entanglement assisted code for a compound quantum channel $\left\{\cN^{(i)}_{ A \to B}\right\}_{i=1}^s$ in the case of informed sender consists of 
\begin{itemize}
\item An encoding operation $\mathcal{E}_i: ME_A \rightarrow A $ for Alice (notice the dependence of the encoding function on the index $i$).
\item A decoding operation $\mathcal{D} : B E_B\rightarrow M'$ for Bob, with $M'\equiv M$ being the output register such that for all $m$ and for all $i \in [1:s]$,
\beq
\Pr\left\{M'\neq m|M=m\right\} \leq \eps. \nonumber
\enq
\end{itemize}
\end{definition}
We prove the following achievability result: 
\begin{theorem}
\label{achievabilityinformed}
Let $\left\{\cN^{(i)}_{A \to B }\right\}_{i=1}^s$ be a compound quantum channel and let $\eps, \eta \in (0,1)$. Let $A'\equiv A$ be a purifying register. Then for any $R$ satisfying 
\beq
\label{eq:informedptop}
R \leq  \min_{i \in [1:s]} \left(\max _{\ketbra{\psi}_{A A'}}  \Ihepsh{B}{A'}{\eps}_{\cN^{(i)}_{A \to B } (\ketbra{\psi}_{AA'})} + (\log2s)\log \left(\frac{\eta}{6\log(2s)}\right) + \log\frac{\eps}{4s^2}\right),
\enq
there exists an $(R, \eps+3\eta)$-entanglement assisted code for compound channel $\left\{\cN^{(i)}_{A \to B }\right\}_{i=1}^s$ in the case of informed sender. 
\end{theorem}

\begin{proof}
Fix $R$ as given in Equation \eqref{eq:informedptop}. Introduce the registers $A_1,A_2,\ldots A_{s2^{R}}$, such that $A_i\equiv A$ and $A'_1,A'_2,\ldots A'_{s2^{R}}$ such that $A'_i\equiv A'$.  Further, for every message $m \in [1:2^R],$ Alice and Bob share a band of $s$ entangled quantum states of the following form: 
\begin{equation}
\label{eq:sharedent}
 \ketbra{\psi}^{(1)}_{A_{s(m-1)+1}A'_{s(m-1)+1}}\otimes \ketbra{\psi}^{(2)}_{A_{s(m-1)+2}A'_{s(m-1)+2}},\ldots \ketbra{\psi}^{(s)}_{A_{sm}A'_{sm}},
\end{equation}
where Alice  holds the registers $$A_{s(m-1)+1},A_{s(m-1)+2}, \cdots, A_{sm},$$ and Bob  holds the registers $$A'_{s(m-1)+1},A'_{s(m-1)+2}, \cdots, A'_{sm},$$ and for every $i \in [1:s],$ $\ketbra{\psi}^{(i)}_{AA'} $ is such that it achieves the maximum in $$\max _{\ketbra{\psi}_{A A'}} \Ihepsh{B}{A'}{\eps}_{\cN^{(i)}_{A \to B } (\ketbra{\psi}_{AA'})}.$$ Let
\begin{equation}
\label{convsets}
 S_{A'}:= \conv(\{\psi^{(i)}_{A'}\}_{i\in [1:s]}) \quad  S_{B}:= \conv(\{\cN^{(i)}_{A\to B}(\psi^{(i)}_A)\}_{i\in [1:s]}) .
\end{equation}
Define the following quantum state belonging to the set $S_{A'}$. 
\begin{equation}
\label{eq:avgstate}
\sigma_{A'}:= \frac{1}{s}\sum_{i\in [1:s]}\psi^{(i)}_{A'}
\end{equation}
We observe that for all $i\in [1:s]$, 
\begin{eqnarray}
\label{eq:itildeihat}
\max _{\ketbra{\psi}_{A A'}} \Ihepsh{B}{A'}{\eps}_{\cN^{(i)}_{A \to B } (\ketbra{\psi}_{AA'})} &=& \Ihepsh{B}{A'}{\eps}_{\cN^{(i)}_{A \to B } (\ketbra{\psi}^{(i)}_{AA'})}\nonumber\\ &\leq& \Ihepshres{B}{A'}{\eps}{S_{B}, S_{A'}}_{\cN^{(i)}_{A \to B } (\ketbra{\psi}^{(i)}_{AA'})}\nonumber\\ &\leq& \Ihepstres{B}{A'}{\eps}{S_B}_{\cN^{(i)}_{A \to B } (\ketbra{\psi}^{(i)}_{AA'}), \sigma_{A'}}.
\end{eqnarray}
 For $i \in [1:s]$, let $ 0 \preceq M^{(i)}_{BA'} \preceq \mathbb{I}$ be such that for all $j\in [1:s]$, we have
\beq
\label{optimalmeasurementi}
\Ihepst{B}{A'}{\eps, S_B}_{\cN^{(i)}_{A \to B } (\ketbra{\psi}^{(i)}_{AA'}),\sigma_{A'}}\leq - \log \tr \left[M^{(i)}_{BA'}\left( \cN^{(j)}_{A\to B}(\psi^{(j)}_{A}) \otimes \sigma_{A'}\right) \right ]. \nonumber
\enq
The existence of such an $M^{(i)}_{BA'}$ is guaranteed by Lemma \ref{minimaxoperator2}. Further, as guaranteed by the Neumark's  theorem (Fact \ref{Neumark}), $\forall i \in [1:s]$, let $\Pi^{(i)}_{BA'P}$ be such that $\forall j \in [1:s]$ 
\begin{equation}
\label{eq:productstates}
 \tr \left[M^{(i)}_{BA'} \left( \cN^{(j)}_{A\to B}(\psi^{(j)}_{A}) \otimes \sigma_{A'}\right) \right ] = \tr \left[\Pi^{(i)}_{BA'P} \left( \cN^{(j)}_{A\to B}(\psi^{(j)}_{A}) \otimes \sigma_{A'} \otimes \ketbra{0}_P\right) \right ] ,
\end{equation}
 and  $$ \tr \left[M^{(i)}_{BA'} \left(\cN^{(i)}_{A\to B}(\ketbra{\psi}^{(i)}_{AA'})\right) \right ] = \tr \left[\Pi^{(i)}_{BA'P} \left(\cN^{(i)}_{A\to B}(\ketbra{\psi}^{(i)}_{AA'}) \otimes  \ketbra{0}_P\right) \right ]. $$ Let $\Pi^\star_{BA'P}$  be the operator obtained by setting $$ \frac{\eta}{3\log(2s)}\leftarrow \delta, \quad \cN^{(i)}_{A\to B}(\ketbra{\psi}^{(i)}_{AA'}) \otimes  \ketbra{0}_P \leftarrow \rho^{(i)}, \quad \Pi^{(i)}_{BA'P} \leftarrow \Pi^{(i)}$$ in Lemma \ref{uni}.

Our protocol is as follows:

\vspace{0.1in}

\noindent {\bf{Encoding:}} Alice  on receiving the message $m \in [1:2^R]$ and getting informed that the channel $\cN^{(i)}_{A \to B}$ will be used for transmission sends chooses the register $A_{s(m-1)+i}$ from the band corresponding to the message $m$ and transmits it over the channel. Let ${{\Theta}}^{(m,i)}_{ BA'_1\cdots A'_{s2^{R}}}$ be the quantum state on Bob's registers after Alice's transmission over the channel. Notice that the quantum state ${\Theta}^{(m,i)}_{BA^\prime_k}$ between the register $A'_k$ and the channel output $B$ is the following  
\begin{equation*}
\label{jointstateinformed}
{{\Theta}}^{(m,i)}_{BA^\prime_k}   =
\begin{cases}
\cN^{(i)}_{A \to B} (\ketbra{\psi}^{(i)}_{AA'}) & \mbox{if }~ k =s(m-1)+i; \\
\cN_{A \to B}^{(i)}(\psi^{(i)}_{A}) \otimes \psi^{(k \text{ mod } s)}_{A'}        & \mbox{otherwise,}
\end{cases}
\end{equation*}
where $k \text{ mod } s$ is interpreted as $s$ instead of $0$, when $k$ is a multiple of $s$.

\vspace{0.1in}

\noindent{\bf{Decoding:}} For each $k \in [1:s2^R]$ consider the following operator
\begin{equation}
\Lambda(k):= \mathbb{I}_{A'_1} \otimes \mathbb{I}_{A'_2} \otimes \cdots \Pi^\star_{BA'_{k}P} \otimes \cdots \otimes \mathbb{I}_{A'_{s2^R}}, \nonumber
\end{equation}
where $\Pi^\star_{BA'_{k}P}$ is as discussed above. The decoding POVM element corresponding to $m $ is:
\begin{equation*}
\Omega(m) := \left(\sum_{ k^\prime \in [1:s2^{R}]} \Lambda({k'})\right)^{-\frac{1}{2}}\left(\sum_{k\in [s(m-1)+1: sm]}\Lambda({k})\right)\left(\sum_{ k^\prime \in [1:s2^{R}]} \Lambda({k'})\right)^{-\frac{1}{2}}. 
\end{equation*}
It is easy to observe that $\sum_m \Omega(m) \preceq \mathbb{I}$. Bob on receiving the channel output appends an ancilla $ \ketbra{0}_P$ to his registers and then measures his registers using the POVM defined above. He outputs `$0$' for the outcome corresponding to the POVM element $\id - \sum_m \Omega(m)$.

\vspace{2mm}

\noindent {\bf{Probability of error:}} Let $M$ be the message which was transmitted by Alice  using the strategy above and let $M'$ be the decoded message by Bob  using the above mentioned decoding POVM. Further, let us assume that the channel $\cN^{(i)}_{ A \to B}$ is used for this transmission. By the symmetry of the encoding and decoding strategy, it is enough to show that $\Pr \left\{M' \neq 1 \mid  M=1 \right\} \leq \eps+ 3\eta$, under the event that $M=1$ is the transmitted message. 
\begin{align}
\label{eq:informedbetter}
& \Pr \left\{M' \neq 1 | M=1\right\} \nonumber\\ & = \tr \left[\left(\mathbb{I} -\Omega(1)\right){\Theta}^{(1,i)}_{B A^\prime_1A^\prime_1\cdots A^\prime_{s2^{R}}} \otimes  \ketbra{0}_P\right]\nonumber\\
& \overset{a} \leq   \frac{\eps+2\eta}{\eps+\eta}\cdot\tr \left[\left(\mathbb{I} -\Lambda(i)\right){\Theta}^{(1,i)}_{B A^\prime_1A^\prime_1\cdots A^\prime_{s2^{R}}}\otimes \ketbra{0}_P\right] + (4+\frac{\eta}{\eps})\sum_{m'\neq 1}\sum_{k\in [m'(s-1)+1:sm']} \Tr \left[\Lambda(k){\Theta}^{(1,i)}_{B A^\prime_1A^\prime_1\cdots A^\prime_{s2^{R}}} \otimes  \ketbra{0}_P \right]\nonumber\\
& \overset{b} =   \frac{\eps+2\eta}{\eps+\eta}\cdot\tr \left[\left(\mathbb{I} - \Pi^\star_{BA'_iP}\right){\Theta}^{(1,i)}_{B A^\prime_i}\otimes  \ketbra{0}_P\right] + (4+\frac{\eta}{\eps})\sum_{m'\neq 1}\sum_{k\in  [m'(s-1)+1:sm']} \Tr \left[\Pi^\star_{BA'_{k}P}{\Theta}^{(1,i)}_{B A^\prime _{k}}\otimes  \ketbra{0}_P \right]\nonumber\\
& = \frac{\eps+2\eta}{\eps+\eta}\cdot\tr \left[\left(\mathbb{I} - \Pi^\star_{BA'_iP}\right){\Theta}^{(1,i)}_{B A^\prime_i}\otimes  \ketbra{0}_P\right] + (4+\frac{\eta}{\eps})\sum_{m'\neq 1}\sum_{j\in  [1:s]} \Tr \left[\Pi^\star_{BA'P}\cN_{A \to B}^{(i)}(\psi^{(i)}_{A}) \otimes \psi^{(j)}_{A'}\otimes  \ketbra{0}_P \right]\nonumber\\
&= \frac{\eps+2\eta}{\eps+\eta}\cdot\tr \left[\left(\mathbb{I} - \Pi^\star_{BA'_iP}\right){\Theta}^{(1,i)}_{B A^\prime_i}\otimes  \ketbra{0}_P\right] + \frac{4\eta}{\eps}s\sum_{m'\neq 1}\Tr \left[\Pi^\star_{BA'P}\cN_{A \to B}^{(i)}(\psi^{(i)}_{A}) \otimes \left(\frac{1}{s}\sum_{j\in [1:s]}\psi^{(j)}_{A'}\right)\otimes  \ketbra{0}_P \right]\nonumber\\
&= \frac{\eps+2\eta}{\eps+\eta}\cdot\tr \left[\left(\mathbb{I} - \Pi^\star_{BA'_iP}\right){\Theta}^{(1,i)}_{B A^\prime_i}\otimes  \ketbra{0}_P\right] + \frac{4\eta}{\eps}s\sum_{m'\neq 1}\Tr \left[\Pi^\star_{BA'P}\cN_{A \to B}^{(i)}(\psi^{(i)}_{A}) \otimes \sigma_{A'}\otimes  \ketbra{0}_P \right]\nonumber\\
&\overset{c} \leq \frac{\eps+2\eta}{\eps+\eta} (\eps + \eta) + \frac{4\eta}{\eps}s^2\times \left(\frac{6\log(2s)} {\eta}\right)^{\log(2s)} \times  \text{exp}\left(\ln 2\cdot \left(R - \min_{j\in [1:s]}\Ihepstres{B}{A'}{\eps}{S_B}_{\cN^{(j)}_{A \to B } (\ketbra{\psi}^{(j)}_{AA'}),\sigma_{A'}}\right)\right)\nonumber\\
&\overset{d}\leq \eps+2\eta + \frac{4\eta}{\eps}s^2\times \left(\frac{6\log(2s)} {\eta}\right)^{\log(2s)} \times  \text{exp}\left(\ln 2\cdot \left(R - \min_{j\in [1:s]}\Ihepsh{B}{A'}{\eps}_{\cN^{(j)}_{A \to B } (\ketbra{\psi}^{(j)}_{AA'})}\right)\right)\nonumber\\
& \overset{e}\leq \eps+3\eta.
\end{align}
where $a$ follows from Hayashi-Nagaoka operator inequality (Fact \ref{haynag}) with $c=\frac{\eta}{\eps+\eta}$ and the identity $ \Lambda(i)\preceq\sum_{k\in [1:s]}\Lambda(k)$; $b$ follows from the definition of $\Lambda (m)$; $c$ follows from the properties of $\Pi^\star_{BA'P}$ (see Lemma \ref{uni}); $d$ follows from Equation \ref{eq:itildeihat} and $e$ follows from our choice of $R.$

\end{proof}

\noindent{\bf Remark:} Observe that from inequality (c) in Equation \ref{eq:informedbetter} and Equation \ref{eq:itildeihat}, the amount of achievable communication is larger than that given in the statement of Theorem \ref{achievabilityinformed}. That is, we have the following corollary.

\begin{corollary}
\label{cor:informed}
Let $\left\{\cN^{(i)}_{A \to B }\right\}_{i=1}^s$ be a compound quantum channel and let $\eps, \eta \in (0,1)$. Let $A'\equiv A$ be a purifying register. Fix the quantum states $\{\ketbra{\psi}^{(i)}_{AA'}\}_{i=1}^s$. Then for any $R$ satisfying 
\begin{equation*}
R \leq  \min_{i \in [1:s]} \left(\Ihepshres{B}{A'}{\eps}{S_{B}, S_{A'}}_{\cN^{(i)}_{A \to B } (\ketbra{\psi}^{(i)}_{AA'})} + (\log2s)\log \left(\frac{\eta}{6\log(2s)}\right) + \log\frac{\eps}{4s^2}\right),
\end{equation*}
where $$S_{A'}:= \conv(\{\psi^{(i)}_{A'}\}_{i\in [1:s]}) \quad  S_{B}:= \conv(\{\cN^{(i)}_{A\to B}(\psi^{(i)}_A)\}_{i\in [1:s]}),$$  there exists an $(R, \eps+3\eta)$-entanglement assisted code for compound channel $\left\{\cN^{(i)}_{A \to B }\right\}_{i=1}^s$ in the case of informed sender. 
\end{corollary}
Above corollary shall help us in deriving the asymptotic result below. For the simplicity of presentation, we have given the achievability in Theorem \ref{achievabilityinformed} in terms of $\Ihepsh{B}{A'}{\eps}_{\cN^{(i)}_{A \to B } (\ketbra{\psi}_{AA'})}$.

\subsection*{Asymptotic and i.i.d. analysis for informed sender}

Now, we proceed to the asymptotic and i.i.d. analysis of our achievability bound in Theorem \ref{achievabilityinformed}. This recovers the result obtained in \cite{BertaGW17}. It can be noted that this rate cannot be exceeded, as it is a minimum over the capacity of all the channels in the given set. 
\begin{theorem}
\label{theo:asympinfo}
Let $\left\{\cN^{(i)}_{A \to B }\right\}_{i=1}^s$ be a compound quantum channel. Let $A'\equiv A$ be a purifying register. Let $C$ satisfy 
\begin{equation*}
C \leq \min_{i\in [1:s]} \left(\max_{\ketbra{\psi}_{AA'}}\mutinf{B}{A'}_{\cN^{(i)}_{A\to B}(\ketbra{\psi}_{AA'})}\right).
\end{equation*}
For every $\eps \in (0,1), \delta>0$, there exists a large enough $n$ such that there exists a $(n(C - \delta), 2\eps)$-entanglement assisted code for the compound quantum channel $\left\{\cN^{(i) \otimes n}_{A \to B }\right\}_{i=1}^s$, in the case of the informed sender.
\end{theorem}
\begin{proof}
 Let $\ketbra{\psi}^{(i)}_{AA'}$ be the quantum state that achieves the supremum in 
$$\max_{\ketbra{\psi}_{AA'}}\mutinf{B}{A'}_{\cN^{(i)}_{A\to B}(\ketbra{\psi}_{AA'})}.$$
Define 
 $$S_{A'^n}:= \conv(\{\psi^{(i)\otimes n}_{A'}\}_{i\in [1:s]}) \quad  S_{B^n}:= \conv(\{\cN^{(i)}_{A\to B}(\psi^{(i)}_A)^{\otimes n}\}_{i\in [1:s]}) .$$
We now apply Lemma \ref{informedasymptote} to Corollary \ref{cor:informed} for the channel $\left\{\cN^{(i) \otimes n}_{A \to B }\right\}_{i=1}^s$ and the sets $S_{A'^n}, S_{B^n}$. The proof follows by choosing $n$ large enough such that $\delta \geq O(n^{-\frac{1}{3}})$. 
\end{proof}
The following lemma is used in above theorem.
\begin{lemma}
\label{informedasymptote}
For all $i\in [1:s]$, it holds that 
\begin{eqnarray*}
\Ihepshres{B^n}{A'^n}{\eps}{S_{B^n}, S_{A'^n}}_{\cN^{(i)}_{A\to B}(\ketbra{\psi}^{(i)}_{AA'})^{\otimes n}} &\geq& n\cdot\relent{\cN^{(i)}_{A\to B}(\ketbra{\psi}^{(i)}_{AA'})}{\cN^{(i)}_{A\to B}(\psi^{(i)}_{A})\otimes \psi^{(i)}_{A'}} - O(n^{\frac{2}{3}})\\ &=& n\cdot\mutinf{B}{A'}_{\cN^{(i)}_{A\to B}(\ketbra{\psi}^{(i)}_{AA'})} - O(n^{\frac{2}{3}}). 
\end{eqnarray*}
\end{lemma}
\begin{proof}
For brevity, we set $\rho^{(i)}_{BA'} : = \cN^{(i)}_{A\to B}(\ketbra{\psi}^{(i)}_{AA'})$. In this terminology, $S_{B^n}$ is the convex hull of the states $\{\rho^{(i) \otimes n}_B\}_{i\in [1:s]}$. Assume that $\ell := \lceil n^{\frac{1}{3}}\rceil$ divides $n$ (without loss of generality). Define the following quantum states
$$\mu_{A'^\ell} := \frac{1}{s}\sum_{i\in [1:s]} \rho^{(i)\otimes \ell}_{A'}, \quad \omega_{B^\ell} := \frac{1}{s}\sum_{i\in [1:s]} \rho^{(i)\otimes \ell}_{B}.$$
Our proof requires the following claims, which are proved towards the end. 
\begin{claim}
\label{clm:dmaxbound}
Let $\sigma_{A'^n}\in S_{A'^n}$ be a quantum state. Then 
$$\sigma_{A'^n} \preceq s^{\frac{n}{\ell}}\left( \mu_{A'^\ell}\right)^{\otimes \frac{n}{\ell}}.$$
Similarly, let $\tau_{B^n}\in S_{B^n}$ be a quantum state. Then  
$$\tau_{B^n} \preceq s^{\frac{n}{\ell}}\left( \omega_{B^\ell}\right)^{\otimes \frac{n}{\ell}}.$$
\end{claim}
\begin{claim}
\label{clm:varentupper}
For each $i\in [1:s]$, it holds that 
$$\varent{\rho_{BA'}^{{(i)}\otimes \ell}}{\omega_{B^\ell}\otimes\mu_{A'^\ell}}\leq \left(2\log s +  \ell\cdot\imax(B:A')_{\rho^{(i)}_{BA'}}\right)^2.$$
\end{claim}
\noindent Assuming the claims, we fix a pair of quantum states $\tau_{B^n}\in S_{B^n}$ and $\sigma_{A'^n}\in S_{A'^n}$. Consider,
\begin{eqnarray*}
\dheps{\rho_{BA'}^{(i)\otimes n}}{\tau_{B^n}\otimes \sigma_{A'^n}}{\eps} &\overset{a}\geq& \dheps{\rho_{BA'}^{(i)\otimes n}}{\left( \omega_{B^\ell}\right)^{\otimes \frac{n}{\ell}}\otimes \left( \mu_{A'^\ell}\right)^{\otimes \frac{n}{\ell}}}{\eps} - \frac{n}{\ell}\log s \\ &=&  \dheps{\left(\rho_{BA'}^{(i)\otimes \ell}\right)^{\otimes \frac{n}{\ell}}}{\left( \omega_{B^\ell}\otimes  \mu_{A'^\ell}\right)^{\otimes \frac{n}{\ell}}}{\eps} - \frac{n}{\ell}\log s \\ &\overset{b}\geq& \frac{n}{\ell}\relent{\rho_{BA'}^{(i)\otimes \ell}}{ \omega_{B^\ell}\otimes  \mu_{A'^\ell}}+ O(\log n) - \frac{n}{\ell}\log s\\& - & \left(\frac{n}{\ell}\cdot l^2\cdot (\log s + \imax(B:A')_{\rho^{(i)}_{BA'}})\right)^{\frac{1}{2}}|\Phi^{-1}(\eps)| \\&\overset{c}\geq& n\relent{\rho_{BA'}^{(i)}}{\rho^{(i)}_B\otimes \rho^{(i)}_{A'}}+ O(\log n) - \frac{n}{\ell}\log s\\& - & \left(n\ell\cdot (\log s + \imax(B:A')_{\rho^{(i)}_{BA'}})\right)^{\frac{1}{2}}|\Phi^{-1}(\eps)|\\ &\overset{d}=& n\relent{\rho_{BA'}^{(i)}}{\rho^{(i)}_B\otimes \rho^{(i)}_{A'}} - O(n^{\frac{2}{3}}),
\end{eqnarray*} 
where $a$ follows from Fact \ref{dhdmax} and Claim \ref{clm:dmaxbound}; $b$ follows from \cite[Equation34]{TomHay13}(also \cite{li2014}) and Claim \ref{clm:varentupper}; $c$ follows from Fact \ref{marginalbest} and $d$ follows from the choice of $\ell$.  This proves the lemma by minimizing over all $\tau_{B^n}\in S_{B^n}$ and $\sigma_{A'^n}\in S_{A'^n}$.

\vspace{0.1in}

\noindent {\bf Proof of Claim \ref{clm:dmaxbound}:} We prove the statement for $\sigma_{A'^n}\in S_{A'^n}$. The statement for $\tau_{B^n}\in S_{B^n}$ follows similarly. Let $\lambda_i$ be such that $\sigma_{A'^n} = \sum_{i\in [1:s]}\lambda_i \rho^{(i)\otimes n}_{A'}$. Since $\lambda_i\leq 1$, we have 
$$\sum_{i\in [1:s]}\lambda_i \rho^{(i)\otimes n}_{A'}\preceq \sum_{i\in [1:s]} \rho^{(i)\otimes n}_{A'} \preceq \left(\sum_{i\in [1:s]}\rho^{(i)\otimes \ell}_{A'}\right)^{\otimes \frac{n}{\ell}} = s^{\frac{n}{\ell}}(\mu_{A'_\ell})^{\otimes \frac{n}{\ell}}.$$ This completes the proof.

\vspace{0.1in}

\noindent {\bf Proof of Claim \ref{clm:varentupper}:} Using Fact \ref{varentmaxent}, it suffices to show the following. 
$$\rho^{(i)\otimes\ell}_{BA'} \preceq 2^{\ell\cdot\imax(B:A')_{\rho^{(i)}_{BA'}}}\rho^{(i)\otimes \ell}_B\otimes \rho^{(i)\otimes \ell}_{A'} = 2^{\ell\cdot\imax(B:A')_{\rho^{(i)}_{BA'}}}\cdot s^2\cdot \frac{1}{s}\rho^{(i)\otimes \ell}_B\otimes \frac{1}{s}\rho^{(i)\otimes \ell}_{A'}$$ $$\preceq 2^{\ell\cdot\imax(B:A')_{\rho^{(i)}_{BA'}}}\cdot s^2\cdot \omega_{B^\ell}\otimes \mu_{A'^\ell}.$$
This completes the proof.

\end{proof}

\section{Application: composite quantum hypothesis testing}
\label{sec:composehypo}

We define the problem of composite quantum hypothesis testing, as introduced in \cite{BertaBH17}.
\begin{definition}
\label{def:composehypo}
Fix a Hilbert space $\cH$ and let $S_1, S_2\subseteq \cD(\cH)$. For an integer $n$ and $i \in \{1,2\}$, let $S^n_i \defeq \text{conv}(\{\rho^{\otimes n}: \rho \in S_i\})$. For an $\eps \in (0,1)$, define 
$$\beta(n, \eps) := \max_{0 \preceq \Lambda \preceq \id: \forall \rho_n\in S^n_1, \Tr(\rho_n\Lambda)\geq 1-\eps} \left(\min_{\sigma_n\in S^n_2} \log\frac{1}{\Tr(\sigma_n \Lambda)}\right).$$
\end{definition}

We show the following, reproducing the main result of \cite{BertaBH17} when $S_2$ is a finite set.
\begin{theorem}
\label{theo:composehypo}
Fix sets $S_1, S_2 \subseteq \cD(\cH)$ and let $\eps, \delta \in (0,1)$. Then
\begin{eqnarray}
\lim_{n\rightarrow \infty} \frac{1}{n}\beta(n, \eps + 2\delta) &\geq & \lim_{n\rightarrow \infty}  \frac{1}{n}\min_{\rho\in S_1}\min_{\sigma_n \in S^n_2}\dmineps{\rho^{\otimes n}}{\sigma_n}{\eps - 2\delta}\label{eq:composehypoeq1}\\
&\geq & \lim_{n\rightarrow \infty}\frac{1}{n}\min_{\rho\in S_1}\relent{\rho^{\otimes n}}{\frac{1}{|S_2|}\sum_{\sigma\in S_2} \sigma^{\otimes n}}, \label{eq:composehypoeq2}
\end{eqnarray}
where Equation \ref{eq:composehypoeq2} holds if $S_2$ is a finite set. 
\end{theorem}
\begin{proof}
Fix an integer $n>1$ and $\eps \in (0,1)$. Fix a net $N_{\delta/n}$ over $\cD(\cH)$ such that for every $\rho'\in \cD(\cH)$, there exists a $\rho\in N_{\delta/n}$ with $\F(\rho, \rho')\geq 1-\frac{\delta^2}{n}$. From \cite{Vershynin12}, one can choose $|N_{\delta/n}| \leq (\frac{2n}{\delta^2})^{|\cH|}$. For every $\rho \in S_1$, let $\rho_{\approx}$ be a quantum state from $N_{\delta^2/n}$ such that $\F(\rho, \rho_{\approx}) \geq 1-\frac{\delta^2}{n}$. Let $S'_1$ be the set of all quantum states $\rho_{\approx}$. It holds that $|S'_1|\leq |N_{\delta/n}| \leq (\frac{2n}{\delta})^{|\cH|}$. Define
\begin{equation}
\label{eq:defbbeta}
\bbeta(n, \eps, \rho_\approx) := \min_{\sigma_n \in S^n_2}\dmineps{\rho_{\approx}^{\otimes n}}{\sigma_n}{\eps} \geq \min_{\sigma_n \in S^n_2}\dmineps{\rho^{\otimes n}}{\sigma_n}{\eps - 2\delta},
\end{equation}
 where we use Fact \ref{fact:smoothdh} and the relation $\Pur(\rho^{\otimes n}, \rho_\approx^{\otimes n}) = \sqrt{1- \F^n(\rho, \rho_\approx)} \leq 2\delta$.
Using the minimax theorem (Fact \ref{fact:minimax}), and the fact that $S^n_2$ is a convex set, we have
\begin{eqnarray*}
2^{-\bbeta(n, \eps, \rho_\approx)} &=& \max_{\sigma_n\in S^n_2}\hspace{0.1in}\min_{0 \preceq \Lambda \preceq \id: \Tr(\Lambda\rho_{\approx}^{\otimes n})\geq 1-\eps} \Tr(\Lambda\sigma_n)\\
&=& \min_{0 \preceq \Lambda \preceq \id: \Tr(\Lambda\rho_{\approx}^{\otimes n})\geq 1-\eps}\hspace{0.1in}\max_{\sigma_n\in S^n_2} \Tr(\Lambda\sigma_n).
\end{eqnarray*}
Fix a $\rho_\approx \in S'_1$. Let $\Lambda(\rho_\approx)$ be the resulting operator such that for all $\sigma_n\in S^n_2$, $\Tr(\Lambda(\rho_\approx)\sigma_n) \leq 2^{-\bbeta(n, \eps, \rho_\approx)}$ and $\Tr(\Lambda(\rho_\approx)\rho_{\approx}^{\otimes n}) \geq 1-\eps$. Let $\Pi(\rho_\approx)$ be the projector obtained by applying Neumark's Theorem (Fact \ref{Neumark}), such that 
$$\Tr(\Pi(\rho_\approx)\rho_\approx^{\otimes n}\otimes \ketbra{0}) \geq 1-\eps, \quad \Tr(\Pi(\rho_\approx)\sigma_n\otimes \ketbra{0}) = \Tr(\Lambda(\rho_\approx)\sigma_n)\leq 2^{-\bbeta(n, \eps, \rho_\approx)}.$$
From Lemma \ref{uni}, there exists a projector $\Pi^*$ such that for all $\rho_\approx\in S'_1$, 
$$\Tr(\Pi^*\rho_\approx^{\otimes n}\otimes \ketbra{0}) \geq 1-\eps - \delta$$ and 
for all $\sigma_n \in S^n_2$, 
\begin{eqnarray*}
\Tr(\Pi^*\sigma_n\otimes \ketbra{0}) &\leq& 2^{\log|S'_1| + \log(2|S'_1|)\log\frac{4\log|S'_1|}{\delta}}\cdot 2^{- \min_{\rho_\approx \in S'_1}\bbeta(n, \eps, \rho_\approx)}\\ &\leq& 2^{4\log|S'_1|\log\frac{\log|S'_1|}{\delta}}\cdot 2^{-\min_{\rho_\approx \in S'_1}\bbeta(n, \eps, \rho_\approx)}.
\end{eqnarray*}
Defining $\Lambda^* := \bra{0}\Pi^*\ket{0}$ and using the linearity of trace, we conclude that for all $\rho_n \in S^n_1$ and $\sigma_n\in S^n_2$, we have
$$\Tr(\Lambda^*\rho_n) \geq 1-\eps - 2\delta, \quad \Tr(\Lambda^*\sigma_n) \leq 2^{4\log|S'_1|\log\frac{\log|S'_1|}{\delta}}\cdot 2^{-\min_{\rho_\approx \in S'_1}\bbeta(n, \eps, \rho_\approx)}.$$
Thus, from Definition \ref{def:composehypo}, we obtain
\begin{eqnarray*}
\beta(n, \eps + 2\delta) &\geq& \min_{\rho_\approx \in S'_1}\bbeta(n, \eps, \rho_\approx) - 4\log|S'_1|\log\frac{\log|S'_1|}{\delta}\\
&\overset{a}\geq& \min_{\rho\in S_1}\min_{\sigma_n \in S^n_2}\dmineps{\rho^{\otimes n}}{\sigma_n}{\eps - 2\delta} - 4\log|S'_1|\cdot\log\frac{\log|S'_1|}{\delta}\\
&\geq& \min_{\rho\in S_1}\min_{\sigma_n \in S^n_2}\dmineps{\rho^{\otimes n}}{\sigma_n}{\eps - 2\delta} - \left(4|\cH|\log\frac{2n}{\delta}\right)^2,
\end{eqnarray*}
where $a$ follows from Equation \ref{eq:defbbeta}. Dividing by $n$ and letting $n\rightarrow \infty$, Equation \ref{eq:composehypoeq1} follows. To show Equation \ref{eq:composehypoeq2}, fix $\rho$ and $\sigma_n$ with the associated distribution $p(\sigma)$, such that
$$\sigma_n = \sum_{\sigma \in S_2}p(\sigma) \sigma^{\otimes n} \preceq \sum_{\sigma\in S_2} \sigma^{\otimes n}.$$
Let $\ell$ be an integer to be chosen later. Consider
$$\sigma_n \preceq \sum_{\sigma\in S_2} \sigma^{\otimes n} \preceq \left(\sum_{\sigma\in S_2} \sigma^{\otimes \ell}\right)^{\otimes \frac{n}{\ell}} = |S_2|^{\frac{n}{\ell}}\left(\frac{1}{|S_2|}\sum_{\sigma\in S_2} \sigma^{\otimes \ell}\right)^{\otimes \frac{n}{\ell}}.$$
Hence,  
\begin{eqnarray*}
 \dmineps{\rho^{\otimes n}}{\sigma_n}{\eps - 2\delta} &\overset{a}\geq& \dmineps{\rho^{\otimes n}}{\left(\frac{1}{|S_2|}\sum_{\sigma\in S_2} \sigma^{\otimes \ell}\right)^{\otimes \frac{n}{\ell}}}{\eps - 2\delta} - \frac{n}{\ell}\log|S_2|\\
&\overset{b}\geq& \frac{n}{\ell}\relent{\rho^{\otimes \ell}}{\frac{1}{|S_2|}\sum_{\sigma\in S_2} \sigma^{\otimes \ell}} - \frac{n}{\ell}\log|S_2| \\
&& - \sqrt{\frac{n}{\ell}\varent{\rho^{\otimes \ell}}{\frac{1}{|S_2|}\sum_{\sigma\in S_2} \sigma^{\otimes \ell}}} - O(\log n).
\end{eqnarray*}
where, $a$ follows from Fact \ref{dhdmax} and $b$ follows from \cite[Equation34]{TomHay13}(also \cite{li2014}). Let $\ell$ be chosen such that it satisfies $\frac{1}{\ell}\varent{\rho^{\otimes \ell}}{\frac{1}{|S_2|}\sum_{\sigma\in S_2} \sigma^{\otimes \ell}} \leq \sqrt{n}$. If $\frac{1}{\ell}\varent{\rho^{\otimes \ell}}{\frac{1}{|S_2|}\sum_{\sigma\in S_2} \sigma^{\otimes \ell}} < \ell$, then let $\ell = \sqrt{n}$. As a result, $\ell$ is a monotonically increasing function of $n$, denoted $\ell(n)$. Now,
\begin{eqnarray*}
\lim_{n\rightarrow \infty}\frac{1}{n}\dmineps{\rho^{\otimes n}}{\sigma_n}{\eps - 2\delta} &\geq& \lim_{n\rightarrow \infty}\left(\frac{1}{\ell(n)}\relent{\rho^{\otimes \ell(n)}}{\frac{1}{|S_2|}\sum_{\sigma\in S_2} \sigma^{\otimes \ell(n)}} - \frac{\log|S_2|}{\ell(n)}  - \sqrt{\frac{1}{\sqrt{n}}} - O(\frac{\log n}{n})\right)\\
&=& \lim_{\ell(n)\rightarrow \infty}\left(\frac{1}{\ell(n)}\relent{\rho^{\otimes \ell(n)}}{\frac{1}{|S_2|}\sum_{\sigma\in S_2} \sigma^{\otimes \ell(n)}} - \frac{\log|S_2|}{\ell(n)}\right)\\
&=& \lim_{\ell(n)\rightarrow \infty}\frac{1}{\ell(n)}\relent{\rho^{\otimes \ell(n)}}{\frac{1}{|S_2|}\sum_{\sigma\in S_2} \sigma^{\otimes \ell(n)}},
\end{eqnarray*}
since $|S_2|$ is finite, by assumption. This completes the proof by relabeling $\ell(n)$ with $n$.
\end{proof}

Above theorem allows us to prove the following corollary.

\begin{corollary}
\label{nouniversaltest}
Fix an $\eps\in (0, \frac{1}{3})$. There exists a quantum state $\rho$ and an integer $n$ large enough such that there is no operator $\Lambda_{n, \rho}$ with the following properties. 
\begin{itemize}
\item We have $\Tr(\Lambda_{n, \rho}\rho^{\otimes n}) \geq 1-\eps/2$.
\item  $\Tr(\Lambda_{n,\rho}\sigma^{\otimes n}) \leq 2^{- n(\relent{\rho}{\sigma} - g(n, \eps))}$ for all $\sigma$, where $\lim_{n\rightarrow \infty} g(n, \eps) = 0$.
\end{itemize} 
\end{corollary}   
\begin{proof}
We prove the lemma by contradiction. Fix an $\eps \in (0,\frac{1}{3})$. Suppose for every quantum state $\rho$ and any integer $n$ large enough, there exists an operator $\Lambda_{n, \rho}$ such that $\Tr(\Lambda_{n, \rho}\rho^{\otimes n}) \geq 1-\eps/2$ and $\Tr(\Lambda_{n,\rho}\sigma^{\otimes n}) \leq 2^{- n(\relent{\rho}{\sigma} - g(n, \eps))}$ for all $\sigma$, where $\lim_{n\rightarrow \infty} g(n, \eps) = 0$. Then we have 
$$\min_{\sigma_n \in S^n_2}\dmineps{\rho^{\otimes n}}{\sigma_n}{\eps/2} \geq \min_{\sigma \in S_2}n(\relent{\rho}{\sigma} - g(n, \eps)), $$ since we have $\sigma_n = \int_{\mu(\sigma)} \sigma^{\otimes n}d\mu(\sigma)$. From Equation \ref{eq:composehypoeq1}, we conclude by setting $\delta \leftarrow \frac{\eps}{4}$ that 
$$\lim_{n\rightarrow \infty}\frac{1}{n}\beta(n, 3\eps/2) \geq \min_{\rho\in S_1}\min_{\sigma \in S_2}\relent{\rho}{\sigma}.$$ Here, we do not need to assume that $S_2$ is a finite set, as this assumption is only required for Equation \ref{eq:composehypoeq2}. Thus, $$\lim_{\eps\rightarrow 0}\lim_{n\rightarrow \infty}\frac{1}{n}\beta(n, 3\eps/2) \geq \min_{\rho\in S_1}\min_{\sigma \in S_2}\relent{\rho}{\sigma}.$$
On the other hand, as shown in \cite[Section IV.B]{BertaBH17}, there exist sets $S_1, S_2$ such that 
$$\lim_{\eps\rightarrow 0}\lim_{n\rightarrow \infty}\frac{1}{n}\beta(n, 3\eps/2) < \min_{\rho\in S_1}\min_{\sigma \in S_2}\relent{\rho}{\sigma}.$$
This leads to a contradiction, which completes the proof.
\end{proof}

\subsection*{Conclusion}

We have studied the one-shot entanglement assisted classical capacity of the compound quantum channel (consisting of $s$ channels), in two cases where both the communicating parties are not aware of the channel and where sender is aware of the channel. We have obtained near optimal one-shot bounds for the amount of communication that can be transmitted in the former case, up to an additive factor of $\log s\log\log s$. Our protocol uses the position-based decoding strategy from \cite{AnshuJW17}, for which we develop a new notion of the union of two quantum projectors (different from those considered in \cite{Aaronson06, HarrowLM17}). We also recover the optimal asymptotic and i.i.d. bounds obtained in \cite{BertaGW17} (it is not clear if the bound in \cite{BertaGW17} is near optimal in the one-shot setting). Our strategy for the case of informed sender follows along similar lines. While we are not able to show the one-shot optimality for this case, we can reproduce the optimal asymptotic and i.i.d. results obtained in \cite{BertaGW17}. We leave the problem of obtaining near optimal bounds for this case for future work. As an another application of the union of projectors, we reproduce the composite quantum hypothesis testing result of \cite{BertaBH17}, in the case where the alternate hypothesis is obtained from a finite set. 

\subsection*{Acknowledgment} 

This work is supported by the Singapore Ministry of Education and the National Research Foundation,
through the Tier 3 Grant ``Random numbers from quantum processes'' MOE2012-T3-1-009 and NRF RF Award NRF-NRFF2013-13. 

\bibliographystyle{ieeetr}
\bibliography{References}

\begin{thebibliography}{10}

\bibitem{Shannon}
C.~E. Shannon, ``A mathematical theory of communication,'' {\em The Bell System
  Technical Journal}, vol.~27, pp.~379--423, July 1948.

\bibitem{Teleportation93}
C.~H. Bennett, G.~Brassard, C.~Cr\'epeau, R.~Jozsa, A.~Peres, and W.~K.
  Wootters, ``Teleporting an unknown quantum state via dual classical and
  einstein-podolsky-rosen channels,'' {\em Phys. Rev. Lett.}, vol.~70,
  pp.~1895--1899, Mar 1993.

\bibitem{bennett92}
C.~H. Bennett and S.~J. Wiesner, ``Communication via one- and two-particle
  operators on einstein-podolsky-rosen states,'' {\em Phys. Rev. Lett.},
  vol.~69, no.~20, pp.~2881--2884, 1992.

\bibitem{blackwell1959}
D.~Blackwell, L.~Breiman, and A.~J. Thomasian, ``The capacity of a class of
  channels,'' {\em Ann. Math. Statist.}, vol.~30, pp.~1229--1241, 12 1959.

\bibitem{Wolfowitz59}
J.~{Wolfowitz}, ``{Simultaneous channels},'' {\em Archive for Rational
  Mechanics and Analysis}, vol.~4, pp.~371--386, Jan. 1959.

\bibitem{GamalK12}
A.~E. Gamal and Y.-H. Kim, {\em Network Information Theory}.
\newblock New York, NY, USA: Cambridge University Press, 2012.

\bibitem{BBoche09}
I.~Bjelakovic and H.~Boche, ``Classical capacities of compound and averaged
  quantum channels,'' {\em IEEE Transactions on Information Theory}, vol.~55,
  pp.~3360--3374, July 2009.

\bibitem{Hayashi2009}
M.~Hayashi, ``Universal coding for classical-quantum channel,'' {\em
  Communications in Mathematical Physics}, vol.~289, no.~3, pp.~1087--1098,
  2009.

\bibitem{Bjelakovic2013}
I.~Bjelakovi{\'{c}}, H.~Boche, G.~Jan{\ss}en, and J.~N{\"o}tzel, {\em
  Arbitrarily Varying and Compound Classical-Quantum Channels and a Note on
  Quantum Zero-Error Capacities}, pp.~247--283.
\newblock Berlin, Heidelberg: Springer Berlin Heidelberg, 2013.

\bibitem{Mosonyi15}
M.~Mosonyi, ``Coding theorems for compound problems via quantum {R}\'{e}nyi
  divergences,'' {\em IEEE Transactions on Information Theory}, vol.~61,
  pp.~2997--3012, June 2015.

\bibitem{BBocheN08}
I.~Bjelakovi\'{c}, H.~Boche, and J.~N\"otzel, ``Quantum capacity of a class of
  compound channels,'' {\em Phys. Rev. A}, vol.~78, p.~042331, Oct 2008.

\bibitem{BBocheN09}
I.~Bjelakovic, H.~Boche, and J.~Notzel, ``Entanglement transmission capacity of
  compound channels,'' in {\em 2009 IEEE International Symposium on Information
  Theory}, pp.~1889--1893, June 2009.

\bibitem{Bjelakovic2009}
I.~Bjelakovi{\'{c}}, H.~Boche, and J.~N{\"o}tzel, ``Entanglement transmission
  and generation under channel uncertainty: Universal quantum channel coding,''
  {\em Communications in Mathematical Physics}, vol.~292, no.~1, pp.~55--97,
  2009.

\bibitem{BertaGW17}
M.~Berta, H.~Gharibyan, and M.~Walter, ``Entanglement-assisted capacities of
  compound quantum channels,'' {\em IEEE Transactions on Information Theory},
  vol.~63, pp.~3306--3321, May 2017.

\bibitem{BocheJK2017}
H.~Boche, G.~Jan{\ss}en, and S.~Kaltenstadler, ``Entanglement-assisted
  classical capacities of compound and arbitrarily varying quantum channels,''
  {\em Quantum Information Processing}, vol.~16, p.~88, Feb 2017.

\bibitem{OgawaN00}
T.~Ogawa and H.~Nagaoka, ``Strong converse and stein's lemma in quantum
  hypothesis testing,'' {\em IEEE Transactions on Information Theory}, vol.~46,
  pp.~2428--2433, Nov 2000.

\bibitem{Ogawa:2002}
T.~Ogawa and H.~Nagaoka, ``A new proof of the channel coding theorem via
  hypothesis testing in quantum information theory,'' in {\em Information
  Theory, 2002. Proceedings. 2002 IEEE International Symposium on}, pp.~73--,
  2002.

\bibitem{HyashiN03}
M.~Hayashi and H.~Nagaoka, ``General formulas for capacity of classical-quantum
  channels,'' {\em IEEE Transactions on Information Theory}, vol.~49,
  pp.~1753--1768, July 2003.

\bibitem{Hayashi07}
M.~Hayashi, ``Error exponent in asymmetric quantum hypothesis testing and its
  application to classical-quantum channel coding,'' {\em Phys. Rev. A},
  vol.~76, p.~062301, Dec 2007.

\bibitem{WangR12}
L.~Wang and R.~Renner, ``One-shot classical-quantum capacity and hypothesis
  testing,'' {\em Phys. Rev. Lett.}, vol.~108, p.~200501, May 2012.

\bibitem{MatthewsW14}
W.~Matthews and S.~Wehner, ``Finite blocklength converse bounds for quantum
  channels,'' {\em IEEE Transactions on Information Theory}, vol.~60,
  pp.~7317--7329, Nov 2014.

\bibitem{AnshuJW17}
A.~Anshu, R.~Jain, and N.~A. Warsi, ``One shot entanglement assisted classical
  and quantum communication over noisy quantum channels: A hypothesis testing
  and convex split approach.'' http://arxiv.org/abs/1702.01940, 2017.

\bibitem{Sen12}
P.~Sen, ``Achieving the {H}an-{K}obayashi inner bound for the quantum
  interference channel,'' in {\em 2012 IEEE International Symposium on
  Information Theory Proceedings}, pp.~736--740, July 2012.

\bibitem{Wilde2013}
M.~M. Wilde, ``Sequential decoding of a general classical-quantum channel,''
  {\em Proceedings of the Royal Society of London A: Mathematical, Physical and
  Engineering Sciences}, vol.~469, no.~2157, 2013.

\bibitem{Gao15}
J.~Gao, ``Quantum union bounds for sequential projective measurements,'' {\em
  Phys. Rev. A}, vol.~92, p.~052331, Nov 2015.

\bibitem{Aaronson06}
S.~Aaronson, ``Qma/qpoly /spl sube/ pspace/poly: de-merlinizing quantum
  protocols,'' in {\em 21st Annual IEEE Conference on Computational Complexity
  (CCC'06)}, pp.~13 pp.--273, 2006.

\bibitem{HarrowLM17}
A.~W. Harrow, C.~Y.-Y. Lin, and A.~Montanaro, ``Sequential measurements,
  disturbance and property testing,'' in {\em Proceedings of the Twenty-Eighth
  Annual ACM-SIAM Symposium on Discrete Algorithms}, pp.~1598--1611, 2017.

\bibitem{TomHay13}
M.~Tomamichel and M.~Hayashi, ``A hierarchy of information quantities for
  finite block length analysis of quantum tasks,'' {\em IEEE Transactions on
  Information Theory}, vol.~59, pp.~7693--7710, Nov 2013.

\bibitem{li2014}
K.~Li, ``Second-order asymptotics for quantum hypothesis testing,'' {\em Ann.
  Statist.}, vol.~42, pp.~171--189, 02 2014.

\bibitem{Wolfowitz78}
J.~{Wolfowitz}, {\em Coding Theorems of Information Theory}.
\newblock New York: Springer-Verlag, 1978.

\bibitem{BertaBH17}
M.~Berta, F.~G. S.~L. Brandao, and C.~Hirche, ``On composite quantum hypothesis
  testing.'' https://arxiv.org/abs/1709.07268, 2017.

\bibitem{Hoeffding65}
W.~Hoeffding, ``Asymptotically optimal tests for multinomial distributions,''
  {\em Ann. Math. Statist.}, vol.~36, pp.~369--401, 04 1965.

\bibitem{umegaki1954}
H.~Umegaki, ``Conditional expectation in an operator algebra, i,'' {\em Tohoku
  Math. J. (2)}, vol.~6, no.~2-3, pp.~177--181, 1954.

\bibitem{BuscemiD10}
F.~Buscemi and N.~Datta, ``The quantum capacity of channels with arbitrarily
  correlated noise,'' {\em IEEE Transactions on Information Theory}, vol.~56,
  pp.~1447--1460, 2010.

\bibitem{Datta09}
N.~Datta, ``Min- and max- relative entropies and a new entanglement monotone,''
  {\em IEEE Transactions on Information Theory}, vol.~55, pp.~2816--2826, 2009.

\bibitem{Renner13}
N.~Ciganović, N.~J. Beaudry, and R.~Renner, ``Smooth max-information as
  one-shot generalization for mutual information,'' {\em IEEE Transactions on
  Information Theory}, vol.~60, pp.~1573--1581, 2014.

\bibitem{vonNeumann1928}
J.~V. Neumann, ``Zur theorie der gesellschaftsspiele,'' {\em Math. Annalen.},
  vol.~100, pp.~295--320, 1928.

\bibitem{Jain05}
R.~Jain, ``Distinguishing sets of quantum states.''
  https://arxiv.org/abs/quant-ph/0506205, 2005.

\bibitem{jordan1875}
C.~Jordan, ``Essai sur la g\`{e}om\`{e}trie \'{a} n dimensions,'' {\em Bulletin
  de la S. M. F.}, vol.~3, pp.~103--174, 1875.

\bibitem{Watrouslecturenote}
J.~Watrous, ``Theory of {Q}uantum {I}nformation, lecture notes,'' 2011.
\newblock https://cs.uwaterloo.ca/~watrous/LectureNotes.html.

\bibitem{QiWW17}
H.~Qi, Q.~Wang, and M.~M. Wilde, ``Applications of position-based coding to
  classical communication over quantum channels.''
  https://arxiv.org/abs/1704.01361, 2017.

\bibitem{Vershynin12}
R.~Vershynin, ``Introduction to the non-asymptotic analysis of random
  matrices,'' in {\em Compressed Sensing, Theory and Applications}, Cambridge
  University Press, 2012.

\end{thebibliography}

\end{document}